\documentstyle[11pt,epsf,cite,A4]{article}

\input epsf
\newcommand{\nix}[1]{}
\begin{document}

\begin{center}
{\Large {\bf{Spin Photocurrents in Quantum Wells\\
review part II, (part I: cond-mat/0304266) }}} \normalsize

\vspace{5mm}
{\large {\bf{Sergey D. Ganichev and Wilhelm Prettl }}}\\
\vspace{5mm}
Fakult\"{a}t f\"{u}r Physik,
Universit\"at Regensburg, 93040 Regensburg, Germany,\\
\vspace{5mm}
\subsection*{ABSTRACT}
\end{center}
\vspace{0.2mm}
%
\hspace{0.3in}
%
Spin photocurrents generated by homogeneous optical excitation
with circularly polarized radiation in quantum wells (QWs)  are
reviewed. The absorption of circularly polarized light results in
optical spin orientation due to the transfer of the angular
momentum of photons to electrons of a two-dimensional electron gas
(2DEG). It is shown that  in quantum wells belonging to one of the
gyrotropic crystal classes a non-equilibrium spin polarization of
uniformly distributed electrons causes a directed motion of
electron in the plane of the QW. A characteristic feature of this
electric current, which occurs in unbiased samples, is that it
reverses its direction upon changing the radiation helicity from
left-handed to right-handed and vice versa.

Two microscopic mechanisms are responsible for the occurrence of
an electric current  linked to a uniform spin polarization in a
QW: the spin polarization induced circular photogalvanic effect
and the spin-galvanic effect. In both effects the current flow is
driven by an asymmetric distribution of spin polarized carriers in
{\boldmath$k$}-space of systems with lifted spin degeneracy due to
{\boldmath$k$}-linear terms in the Hamiltonian. Spin photocurrents
provide  methods  to investigate  spin relaxation and to conclude
on the in-plane symmetry of QWs. The effect can also be utilized
to develop fast detectors to determine the degree of circular
polarization of a radiation beam.  Furthermore spin photocurrents
at infrared excitation were used to demonstrate and investigate
monopolar spin orientation of free carriers.
\vspace{15mm}
%
\setcounter{page}{26} \setcounter{subsection}{1}
\setcounter{section}{4} \setcounter{equation}{32}
\setcounter{figure}{15}
\newpage
\tableofcontents
\newpage
\vspace{10mm}
\newpage

\subsection{Spin-galvanic effect}
\label{IVB}

\subsubsection{Spin galvanic effect in the presence of external magnetic field}
\label{IVB1}

\begin{figure}[h]
\centerline{\epsfxsize 86mm \epsfbox{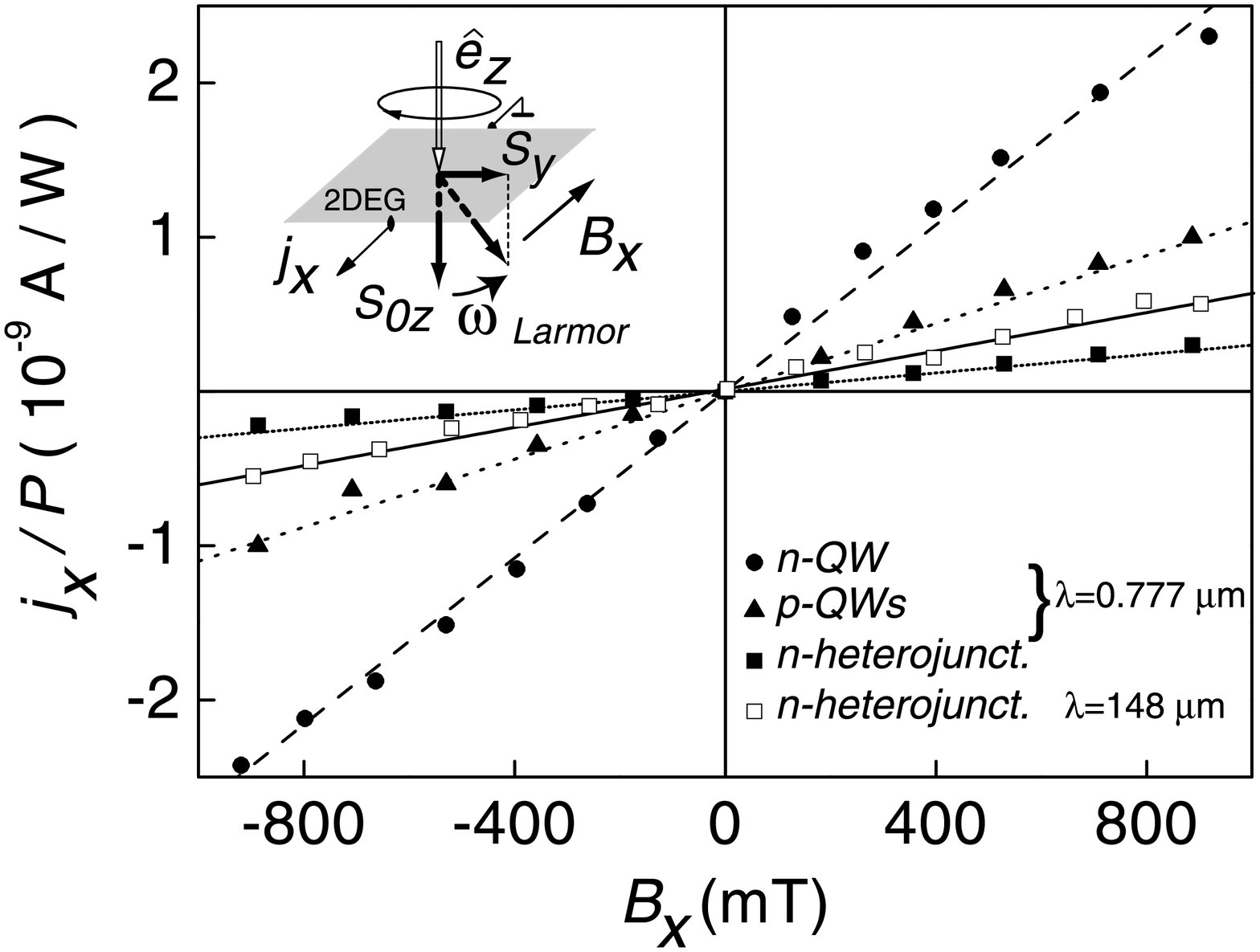  }}
\caption{Spin-galvanic current  $j_x$ normalized by $P$ as a
function of  magnetic field $B$ for  normally incident circularly
polarized radiation at room temperature for various samples and
wavelengths. Full symbols: $\lambda$ = 0.777~$\mu$m, $P$ = 100~mW.
Triangles, squares and circles correspond to $n$-type and $p$-type
multiple QWs, and an $n$-type GaAs/AlGaAs heterojunction,
respectively. Open squares: $n$-type GaAs/AlGaAs heterojunction,
$\lambda = 148~\mu$m, $P$ = 20~kW. The  inset shows   the geometry
of the experiment where $\hat{e}_z$ indicates the direction of the
incoming light. }
\label{fig16}
\end{figure}

\begin{figure}
\centerline{\epsfxsize 86mm \epsfbox{ 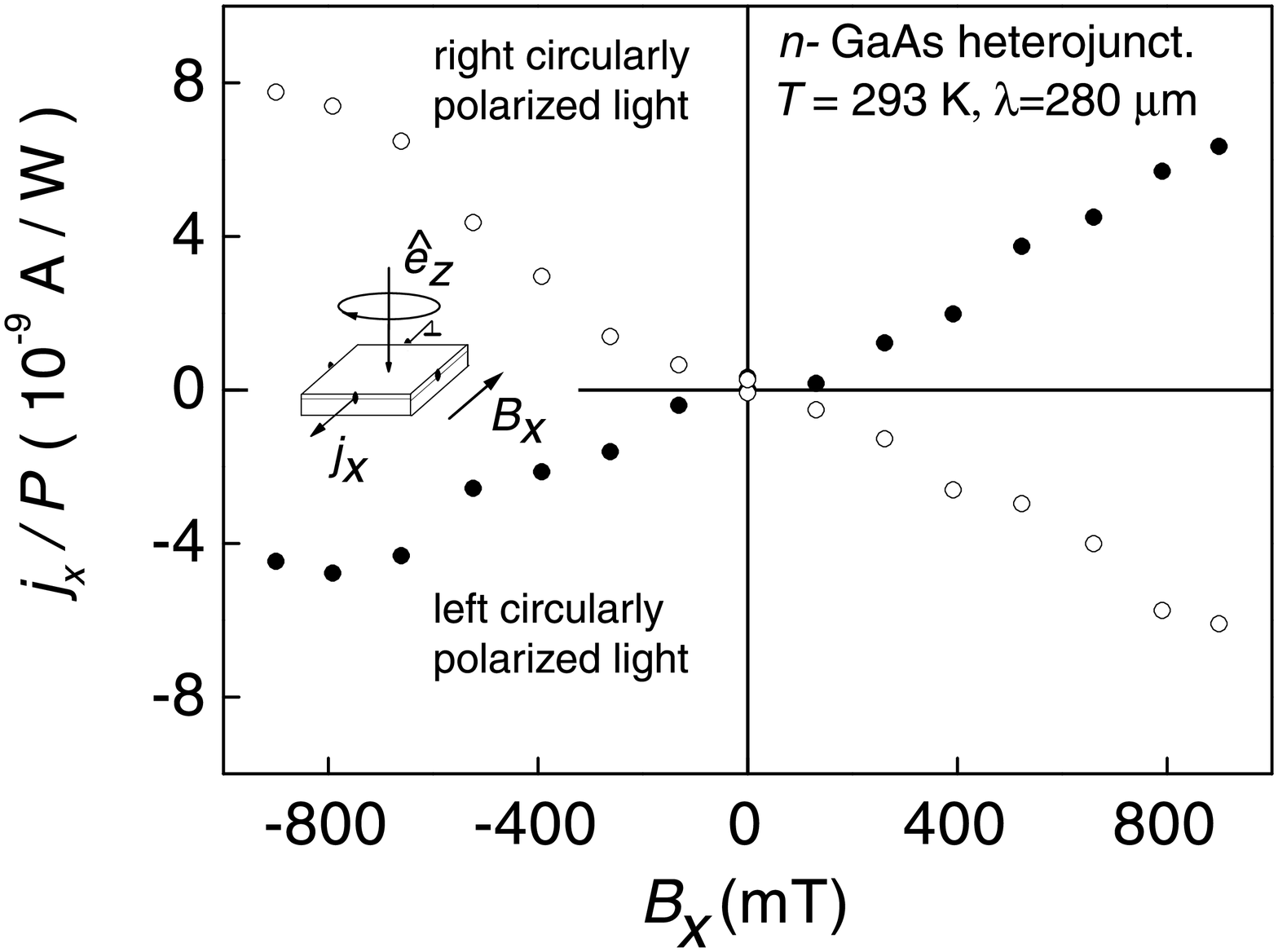 }}
\caption{Magnetic field dependence of the spin-galvanic current
normalized by $P$ achieved by intra-subband transitions within
$e${\it 1} conduction subband by excitation with  radiation of
$\lambda=$280~$\mu$m wavelength. Results are plotted for an
(001)-grown GaAs single heterojunction at room temperature.}
\label{fig17}
\end{figure}

\begin{figure}
\centerline{\epsfxsize 86mm \epsfbox{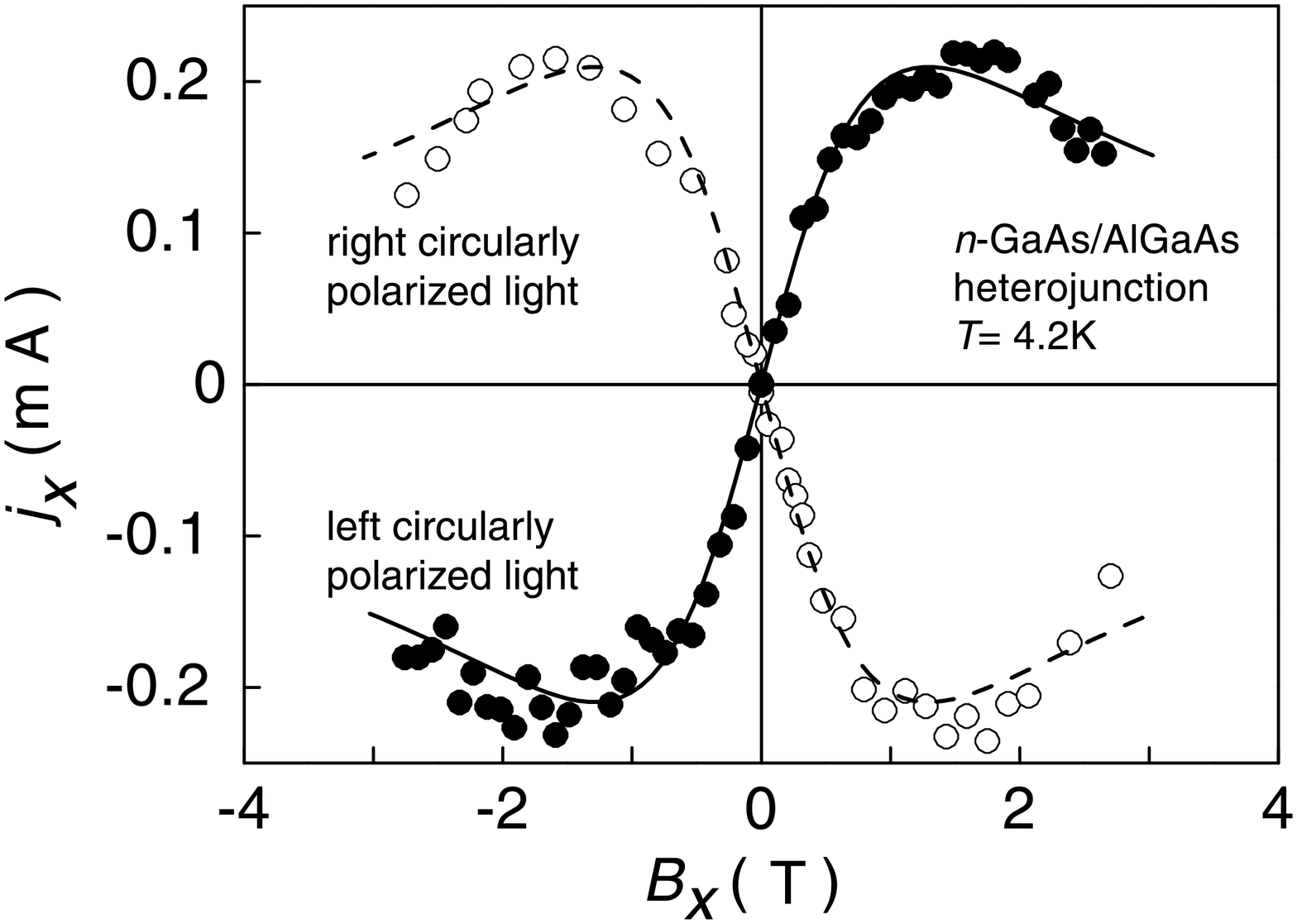  }}
\caption{Spin-galvanic current $j_x$ as a function of magnetic
field $B$ for normally incident right-handed (open circles) and
left-handed (solid circles) circularly polarized radiation at
$\lambda = 148~\mu$m and radiation power 20~kW. Measurements are
presented for an $n$-type GaAs/AlGaAs single heterojunction at
$T=~$4.2~K. Solid and dashed curves are fitted after Eqs.~(25)
and~(30)   using the same value of the spin relaxation time
$\tau_s$  and scaling of the ordinate.}
\label{fig18}
\end{figure}

\begin{figure}[h]
\centerline{\epsfxsize 86mm \epsfbox{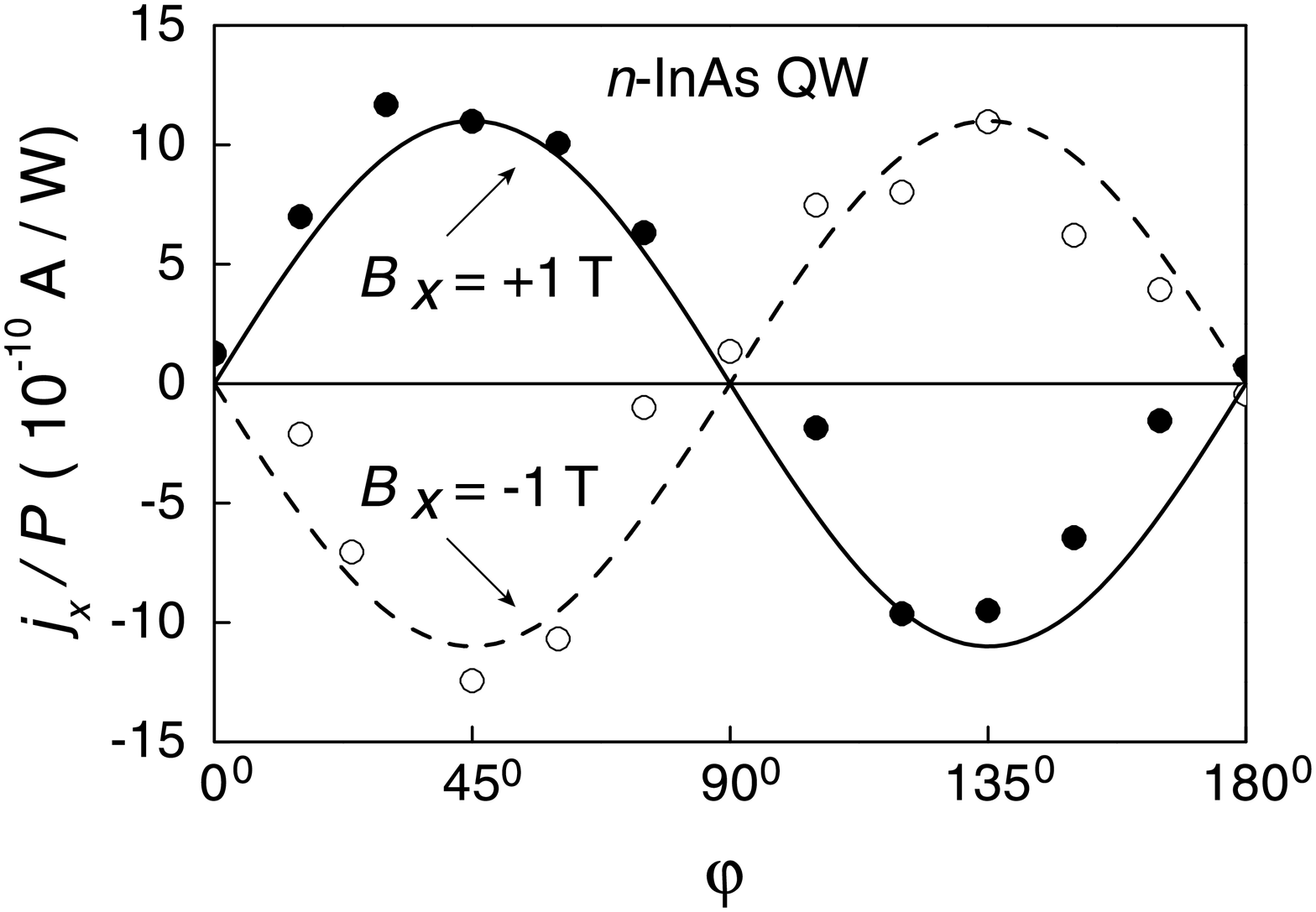  }}
\caption{Spin-galvanic current   normalized by $P$ as a function
of the phase angle $\varphi$ in an (001)-grown $n$-type InAs QW of
15~nm width at $T$~=4.2~K. The photocurrent  excited by normal
incident radiation of $\lambda = 148$~$\mu$m is measured in
$x$-direction parallel (full circles) and anti-parallel (open
circles) to the in-plane magnetic field $B_x$. Solid and dashed
curves are fitted after Eqs.~(32)  using the same scaling of the
ordinate. }
\label{fig19}
\end{figure}

\begin{figure}
\centerline{\epsfxsize 66mm \epsfbox{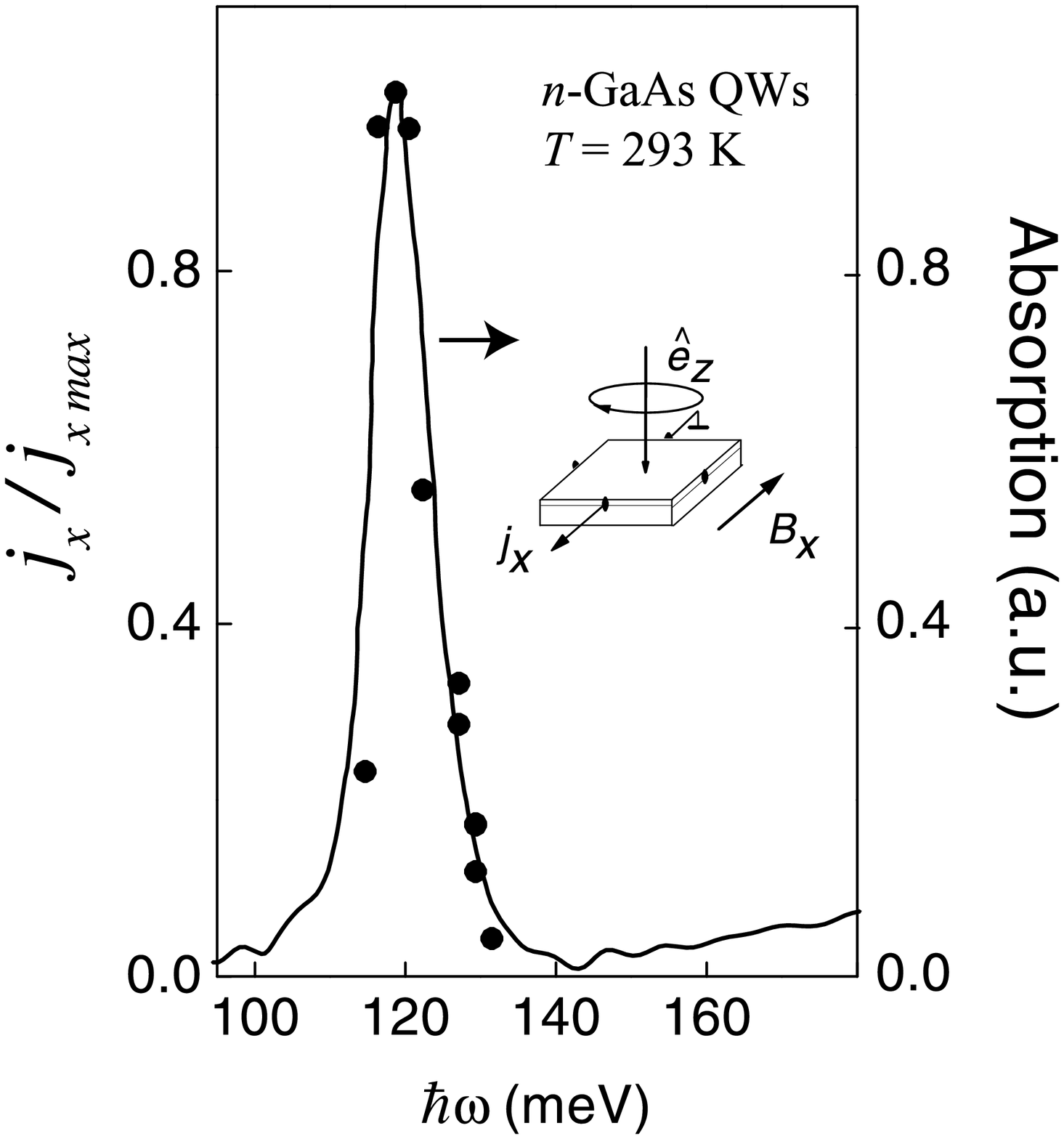  }}
\caption{Spectral dependence of the spin-galvanic effect
(001)-grown $n$-type GaAs QWs of 8.2~nm width at room temperature.
Data (dots)  are presented for optical excitation at normal
incidence of right-handed circularly polarized radiation. A
magnetic field of $B_x= 1$~T was used. For comparison the
absorption spectrum is shown by the full line. }
\label{fig20}
\end{figure}

The spin photocurrent due to the spin-galvanic effect has been
experimentally investigated by the method described in
section~2.3.5 and depicted in
Fig.~7~\cite{Nature02,PASPS02sge,ICPS26,PASPS02monop}. A
homogeneous non-equilibrium spin polarization perpendicular to the
plane of (001)-grown QWs has been prepared by absorption of
circularly polarized radiation at normal incidence. The
measurements were carried out on $n$-type GaAs and InAs samples.
In this experimental configuration the spin polarization does not
yield an electric current. However, applying an in-plane magnetic
field a spin-galvanic current has been observed in $n$-type
materials for both  visible and infrared radiation
(Figs.~\ref{fig16}-\ref{fig20}).

%

For low magnetic fields $B$ where $\omega_L\tau_s < 1$ holds, the
photocurrent  increases linearly as expected from Eqs.~(25)
and~(30). This is seen in the room temperature data of
Figs.~\ref{fig16} and \ref{fig17} as well as in the 4.2~K data in
Fig.~\ref{fig18}  for  $B \leq 1$~T. The polarity of the current
depends on the direction of  the excited spins (see
Figs.~\ref{fig17} and \ref{fig18}, $\pm$ $z$-direction for right
or left  circularly polarized light, respectively) and on the
direction of the applied magnetic field (see
Figs.~\ref{fig16}-\ref{fig19}, $\pm$ $B_x$-direction). For
magnetic field applied along $\langle 110\rangle$ the current is
parallel (anti-parallel) to the magnetic field vector. For $B
\parallel \langle 100\rangle$ both the transverse and the
longitudinal effects are observed~\cite{PASPS02sge}. This
observation as well as helicity dependence of the photocurrent
current shown in Fig.~\ref{fig19} are in good agreement to the
phenomenological relation  (see Eqs.~(32).

%

Comparing  the power sensitivity for visible and infrared
excitation we find them to be of the same order of magnitude as
seen in Fig.~\ref{fig16}.  However, we note that  the current
contribution $per$ $photon$ is by two orders of magnitude larger
for inter-band   excitation compared to intra-subband absorption.
This is  due to a more effective spin generation rate by
inter-band transitions. It may be even larger since the current
gets partially shortened by photogenerated carriers in  the
semi-insulating substrate. For higher magnetic fields the current
assumes a maximum and decreases upon further increase of $B$, as
shown in Fig.~\ref{fig18}. This drop of the current is ascribed to
the Hanle effect~\cite{Meier}. The experimental data are well
described by Eqs.~(25) and (30). The observation of the Hanle
effect demonstrates that free carrier intra-subband transitions
can polarize the spins of electron systems. The measurements allow
to obtain the spin relaxation time $\tau_s$ from the peak position
of the photocurrent where $\omega_L\tau_s = 1$
holds~\cite{Nature02}.

%

In $p$-GaAs QWs at infrared excitation causing spin polarization
of holes only, no spin-galvanic effect could  be
detected~\cite{Nature02,PASPS02sge}. In contrast to infrared
experiments a current signal has been detected in $p$-type samples
for visible excitation which polarize both electrons and holes
(see Fig.~\ref{fig16}). This current is due to the spin
polarization  of electrons only, which are in this case the
minority carriers generated by inter-band excitation. The
spin-galvanic effect in $p$-type material at inter- or
intra-subband excitation could not be observed because of the
experimental procedure which makes use of the Larmor precession to
obtain an in-plane spin polarization. It is due to the fact that
the in-plane $g$-factor for heavy holes is very
small~\cite{Marie99p5811}  which makes the effect of the magnetic
field negligible~\cite{Nature02}. This result does not exclude the
spin-galvanic effect in $p$-type materials which might be
observable by hole injection with spins in the plane of the QW.

Spin photocurrents due to the spin-galvanic effect have been
recorded for inter-band, inter-subband, as well as for
intra-subband transitions~\cite{Nature02,PASPS02sge,PRB03sge}.

%

%

{\em Inter-band transitions} have been investigated in $n$- and
$p$-type GaAs using circularly polarized light of a Ti:sapphire
laser at $\lambda = 0.777~\mu$m. In this experiment electrons are
excited from the valence band to the conduction band yielding a
spin polarization in  the conduction band due to selection rules.

{\em Direct inter-subband transitions} have been achieved in GaAs
QWs of 8.2~nm and 8.6~nm widths   at absorption of radiation in
the range of 9~$\mu$m to 11~$\mu$m
wavelength~\cite{ICPS26,PASPS02monop}. Applying MIR radiation of
the CO$_2$ laser the spin-galvanic current at normal incidence of
radiation  has been observed. In contrast to spin orientation
induced CPGE  the wavelength dependence of the spin-galvanic
effect obtained between 9.2~$\mu$m and 10.6~$\mu$m repeats the
spectral behaviour of direct inter-subband absorption  (see
Fig.~\ref{fig20}). This observation is in agreement with the
mechanism of the spin-galvanic effect and the microscopic theory
presented in section~2.3.3. The occurrence of a spin-galvanic
current requires only a spin polarization in the lower subband and
asymmetric spin relaxation. In the present case the spin
orientation is generated by resonant spin-selective optical
excitation followed by spin-non-specific thermalization. Therefore
the magnitude of the spin polarization and hence the current
depends on the absorption strength but not on the momentum
{\boldmath$k$} of optical transition as in the case of CPGE
described in section~2.2.1.

We would like to emphasize that spin sensitive $e${\it 1}-$e${\it
2} inter-subband transitions in $n$-type QWs have been observed at
normal incidence when there is no component of the electric field
of the radiation normal to the plane of the QWs. Generally it is
believed that inter-subband transitions in $n$-type QWs can only
be excited by infrared light polarized in the growth direction $z$
of the QWs~\cite{book}. Furthermore such transitions are spin
insensitive and, hence, do not lead to optical orientation. Since
the argument, leading to these selection rules, is based on the
effective mass approximation in a single band model, the selection
rules are not rigorous. The mechanism which leads to spin
orientation in this geometry will be discussed in the
section~\ref{IVC1}.

{\em At indirect transitions} the spin-galvanic effect as in the
case of spin orientation induced CPGE has been obtained in
$n$-type GaAs and InAs QWs using FIR radiation (see
Figs.~\ref{fig16}-\ref{fig19}). The presence of the spin-galvanic
effect which is due to spin orientation excitable at MIR and FIR
wavelengths gives clear evidence that direct inter-subband and
Drude absorption of circularly polarized radiations results in
spin orientation. The mechanism of this spin orientation is not
obvious and will be introduced in section~\ref{IVC2}.

\subsubsection{Spin-galvanic effect at optical excitation without external fields}
\label{IVB2}

In the experiments described above an external magnetic field was
used for re-orientation of an optically generated spin
polarization.  The spin-galvanic effect can also be observed at
optical excitation only, without application of an external
magnetic field. The necessary in-plane component of the spin
polarization is obtained by oblique incidence of the exciting
circular polarized radiation. In this case, however, a spin
orientation induced CPGE may also occur interfering with the
spin-galvanic effect. Nevertheless, a pure spin-galvanic current
may be obtained at inter-subband transitions in $n$-type GaAs
QWs~\cite{PRB03sge}. As shown above the spectrum of CPGE changes
sign and vanishes in the center of resonance~\cite{PRB03inv} (see
section~4.1.2 and Eqs.~(22) - (23)). In contrast, the optically
induced spin-galvanic current  is proportional to the absorbance
(Eqs.~(29)) and, hence, assumes a maximum at the center of the
resonance~\cite{ICPS26,PASPS02monop} (see section~\ref{IVB1}).
Thus, if a measurable helicity dependent current is present in the
center of the resonance it must be attributed to the spin-galvanic
effect.

%
\begin{figure}
\centerline{\epsfxsize 66mm \epsfbox{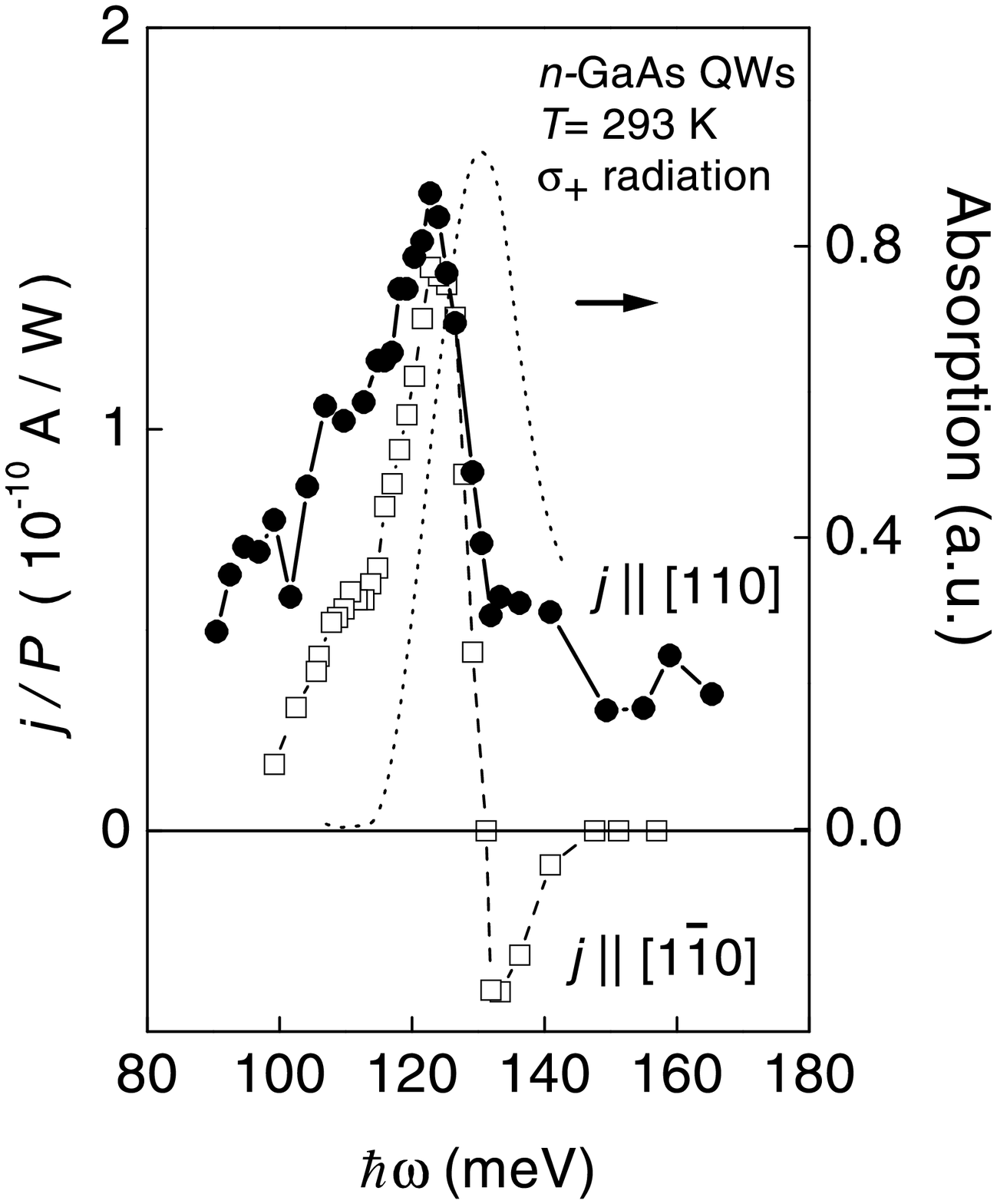}}
\caption{ Photocurrent in QWs normalized by the light power $P$ at
oblique incidence of right-handed circularly polarized radiation
on $n$-type (001)-grown GaAs/AlGaAs QWs of 8.2~nm width at
$T=293$~K as a function of the photon energy $\hbar \omega$.
Circles: current in [110] direction in response to irradiation
parallel [1$\bar{1}$0]. Rectangles:  current in [1$\bar{1}$0]
direction in response to irradiation parallel [110]. The dotted
line shows the absorption measured using a Fourier transform
spectrometer.}
\label{fig21}
\end{figure}

These experiments have been carried out  making use of the
spectral tunability of the free electron laser
''FELIX''~\cite{Knippels99p1578}. The photon energy dependence of
the current was measured for incidence in two different planes
with components of propagation along the $x$- and $y$-directions.
In Fig.~\ref{fig21} the observed current for both directions is
plotted as a function of photon energy $\hbar \omega$ for
$\sigma_+$  radiation together with the absorption spectrum. It
can be seen that for a current along $x$ $\parallel$ [1$\bar{1}$0]
the spectral shape is similar to the derivative of the absorption
spectrum, and in particular there is a change of sign which occurs
at the line center of the absorption. When the sample was rotated
by 90$^\circ$ about $z$ the sign change in the current, now along
$y$ $\parallel$ [110], disappears and its spectral shape follows
more closely the absorption spectrum.

The spectral inversion of sign of the photocurrent in $x$
direction indicates the CPGE which is proportional to  the
derivative of the absorption spectrum (see Eqs.~(22) and (23)). In
contrast with the CPGE the sign of the spin-galvanic current does
not depend on the wavelength (see section~\ref{IVB1}
and~\cite{ICPS26,PASPS02monop}). This can be seen from Fig.~5b
which illustrates the origin of the spin-galvanic effect. All that
is required is a spin orientation of the lower subband, and
asymmetrical spin relaxation then drives a
current~\cite{Nature02}. In our case the spin orientation is
generated by resonant spin-selective optical excitation followed
by spin-non-specific thermalization. The magnitude of the spin
polarization and hence the current depends on the initial
absorption strength but not on the momentum {\boldmath$k$} of
transition. Therefore there is no sign change and the shape of the
spectrum follows the absorption (see Eqs.~(29)
and~\cite{PRB03sge,ICPS26,PASPS02monop}). The lack of a sign
change for current along $y$ $\parallel$ [110]  in the experiment
shows that the spin-galvanic dominates for this orientation.

The non-equivalence   of the  two orientations [110] and
[1$\bar{1}$0] is caused by the interplay of BIA and SIA terms in
the Hamiltonian when rotating the wavevector in the QW plane. Both
currents, CPGE and the spin-galvanic current, are due to  spin
splitting of subbands in  {\boldmath$k$}-space described by
Eq.~(1). The pseudo-tensors {\boldmath$\gamma$}  and
{\boldmath$Q$} determining the current are related to the
transposed pseudo-tensor {\boldmath$\beta$}. They are subjected to
the same symmetry restrictions so that their irreducible
components  differ only by  scalar factors. In $C_{2v}$ symmetry
usually $\beta_{yx} \neq \beta_{xy} $ and it is reasonable to
introduce symmetric and anti-symmetric tensor components
$\beta_{BIA}^{(\nu)} = 1/2(\beta_{xy}^{(\nu)}+\beta_{yx}^{(\nu)})$
and $\beta_{SIA}^{(\nu)} =
1/2(\beta_{xy}^{(\nu)}-\beta_{yx}^{(\nu)})$, where $\nu$=1,2
indicates the $e${\it 1} and $e${\it 2} subbands respectively.
$\beta_{BIA}^{(\nu)} $ and $\beta_{SIA}^{(\nu)}$ result from bulk
inversion asymmetry (BIA) and from structural inversion asymmetry
(SIA), respectively.

As discussed above and sketched in Fig.~5 both CPGE and
spin-galvanic currents, say in $x$  direction, are caused by the
band splitting in $k_x$ direction and therefore are proportional
to $\beta_{yx}$ (for current in $y$-direction one should
interchange the indices $x$ and $y$ ). Then the currents in the
$x$ and $y$ directions read
\begin{equation} \label{equ32a}
j_x =  A_{CPGE} [ (\beta^{(1)}_{BIA} - \beta^{(1)}_{SIA}) -
 (\beta^{(2)}_{BIA} - \beta^{(2)}_{SIA} ) ] P_{circ} \hat{e}_y
+ A_{SGE}( \beta^{(1)}_{BIA} -  \beta^{(1)}_{SIA} ) S_y
\end{equation}
and
\begin{equation} \label{eq32b}
j_y = A_{CPGE} [ (\beta^{(1)}_{BIA} + \beta^{(1)}_{SIA}) -
 (\beta^{(2)}_{BIA} + \beta^{(2)}_{SIA} ) ] P_{circ} \hat{e}_x
+ A_{SGE}( \beta^{(1)}_{BIA} + \beta^{(1)}_{SIA} ) S_x \:\:,
\end{equation}
where $A_{CPGE}$ and $A_{SGE}$ are factors related to
{\boldmath$\gamma$} and {\boldmath$Q$}, respectively, and a
subscript SGE indicates the spin-galvanic effect.

In the present case {\boldmath$S$} is obtained by  optical
orientation, its sign and magnitude are proportional to $P_{circ}$
and it is oriented along the in-plane component of
{\boldmath$\hat{e}$}. The magnitude of  CPGE is determined by the
values of {\boldmath$k$} in the initial and final states, and
hence depends on the spin splitting $\beta_{BIA}$ and
$\beta_{SIA}$ of both $e${\it 1} and $e${\it 2} subbands.  In
contrast, the spin-galvanic effect is due to relaxation between
the spin states of the lowest subband and hence depends only on
$\beta_{BIA}$ and $\beta_{SIA}$ of $e${\it 1}.

The equations above show that in directions $y$ and $x$  the
spin-galvanic effect and the CPGE are proportional to terms with
the sum and the difference respectively of BIA and SIA terms. For
our sample it appears that in the case where they add, the
spin-galvanic effect dominates over CPGE consistent with the lack
of sign change for the current along the $y$-direction in
Fig.~\ref{fig21}. Conversely when BIA and SIA terms subtract the
spin-galvanic effect is suppressed and CPGE dominates. We
emphasize that at the maximum of absorption, where spin
orientation induced CPGE is equal to zero for both directions, the
current obtained is caused solely by the spin-galvanic effect.

\subsection{Monopolar spin orientation}
\label{IVC}

Absorption of circularly polarized light in semiconductors may
result in spin polarization of photoexcited carriers. While this
phenomenon of optical orientation caused by inter-band transitions
in  semiconductors is known since a long
time~\cite{Dyakonov71p144,Lampel68p491,Ekimov70p198,Zakharchenya71p137}
and has been widely studied ~\cite{Meier}, it is not obvious that
free carrier absorption due to inter-subband and intra-subband
transitions can also result in a spin polarization. Observation of
a spin-polarization induced CPGE and the spin-galvanic effect in
the MIR and FIR spectral range unambiguously demonstrates that
spin orientation may be achieved due to free carrier absorption.
This optical orientation may be referred to as
`monopolar'~\cite{PhysicaB99multiphoton} because photon energies
are much less than  the fundamental energy gap and only one type
of carriers, electrons or holes, is excited. Here we consider
mechanisms of monopolar optical orientation due to direct
inter-subband transitions as well as by Drude-like intra-subband
absorption  for $n$- and $p$-type QWs based on zinc-blende
structure semiconductors.

Monopolar spin orientation in $n$-type QWs becomes possible if an
admixture of valence band states to the conduction band wave
function and the spin-orbit splitting of the valence band are
taken into account~\cite{ICPS26,PASPS02monop}. We emphasize that
the spin generation rate under monopolar optical orientation
depends strongly on the energy of spin-orbit splitting of the
valence band, $\Delta_{so}$. It is due to the fact that the
$\Gamma_8$ valence band and the $\Gamma_7$ spin-orbit split-off
band contribute to the matrix element of spin-flip transitions
with  opposite signs. In $p$-type QWs analogous mechanisms are
responsible for spin orientation.

The generation rate of the  electron spin
polarization {\boldmath $S$} due to optical excitation can be written as
\begin{equation}
\dot{S}=s (\eta I/ \hbar \omega) P_{circ} \:, \label{equ34}
\end{equation}
where $s$  is the average electron spin generated per one absorbed
photon of circularly polarized radiation, and $\eta$ is the fraction
of the energy flux absorbed in the QW.

\subsubsection{Direct transitions between size-quantized subbands}
\label{IVC1}

As Eq.~(\ref{equ34}) shows the spin generation rate $\dot{S}$ is
proportional to the absorbance $\eta_{12}$. In order to explain the
observed spin orientation at inter-subband transitions between
$e${\it 1} and $e${\it 2} subbands in $n$-type QWs and, in
particular, the absorption of light polarized in the plane of a QW
the $\mbox{\boldmath$k$} \cdot  \mbox{\boldmath$p$}  $ admixture
of valence band states to the conduction band wave functions has
to be taken into account~\cite{ICPS26,PASPS02monop}. Calculations
yield that inter-subband absorption of circularly polarized light
propagating along $z$ induces only spin-flip transitions resulting
in  100~\% optical orientation of photoexcited carriers, i.e.
$s=1$. In this geometry the fraction of the energy flux absorbed
in the QW by transitions from the first subband $e${\it 1} to the
second subband $e${\it 2} has the form
\begin{equation}
\eta_{12} = \frac{128 \alpha^*}{9 n } \,
\frac{\Delta^2_{so}(2E_g+\Delta_{so})^2 (\varepsilon_{e {\it 2}}-\varepsilon_{e {\it
1}}) \varepsilon_{e {\it 1}}}
{E^2_g(E_g+\Delta_{so})^2(3E_g+2\Delta_{so})^2} \, \frac{\hbar^2
n_s}{m^*_{e{\it 1}} } \,\delta(\hbar\omega - \varepsilon_{e {\it 1}} + \varepsilon_{e {\it
2}}) \: ,
\label{equ35}
\end{equation}
where $\alpha^*$ is the fine structure constant, $n$ is a
refraction index, $n_s$ is a free carrier density, and
$\varepsilon_{e {\it 1}}$ and $\varepsilon_{e {\it 2}}$ are the
energies of  the size-quantized subbands $e${\it 1} and $e${\it
2}, respectively. The $\delta$-function describes the resonant
behaviour of the inter-subband transitions.

In $p$-type QWs, optical orientation is caused by heavy-hole to
light-hole absorption of circularly polarized radiation and occurs for
transitions at in-plane wavevector $\mbox{\boldmath$k$} \neq 0$
due to the mixing of heavy-hole and light-hole
subbands~\cite{Danishevskii85p439}.

\subsubsection{Drude absorption due to indirect intra-subband transitions}
\label{IVC2}

In the far-infrared range where the photon energy is not enough
for direct inter-subband transition in $n$- or  in $p$-type
samples, the absorption of light by free carriers is caused by
indirect intra-subband  transitions where the momentum
conservation law is satisfied due to emission or absorption of
acoustic or optical phonons, static defects etc. (Drude-like
absorption). We assume that the carriers occupy the $e${\it
1}-subband. The intra-subband optical transitions in QWs involving
both the electron-photon interaction and momentum scattering are
described by second-order processes with virtual transitions via
intermediate states. The compound matrix elements for such kind of
transitions with the initial and final states in the same band has
the standard form~\cite{ICPS26}
\begin{equation}
M_{cm'_s{\boldmath k}' \leftarrow cm_s{\boldmath k}}=\sum_{\nu} \left( \frac{
V_{cm'_s{\boldmath k}',\,\nu{\boldmath k}}\,R_{\,\nu{\boldmath k},cm_s{\boldmath k}} }
{\varepsilon_{\nu{\boldmath k}}-\varepsilon_{c{\boldmath k}}-\hbar\omega} + \frac{
R_{\,cm'_s{\boldmath k}',\,\nu{\boldmath k}'}\,V_{\nu{\boldmath k}',cm_s{\boldmath k}} }
{\varepsilon_{\nu{\boldmath k}'}-\varepsilon_{c{\boldmath k}} \pm \hbar\Omega_{{\boldmath k}-{\boldmath k}'}}
\right) \;.
\label{equ36}
\end{equation}
Here $\varepsilon_{c{\boldmath k}}$, $\varepsilon_{c{\boldmath
k}'}$ and $\varepsilon_{\nu}$ are the electron energies in the
initial $|c,m_s, {\boldmath k} \rangle$, final $|c,m'_s,
{\boldmath k}' \rangle$ and intermediate $|\nu \rangle$ states,
respectively, $m_s$ is the spin index, $R$ is the matrix element
of electron interaction with the electromagnetic wave, $V$ is the
matrix element of electron-phonon or electron-defect interaction,
and $\hbar\Omega_{{\boldmath k}-{\boldmath k}'}$ is the energy of
the involved phonon. The sign $\pm$ in Eq.~(\ref{equ36})
correspond to  emission and absorption of phonons. A dominant
contribution to the optical absorption is caused by processes with
intermediate states in the same subband. This is the channel that
determines the coefficient of intra-subband absorbance, $\eta$.
However such transitions conserve the electronic spin and, hence,
do not lead to an optical orientation.

%
\begin{figure}
\centerline{\epsfxsize 120mm \epsfbox{ 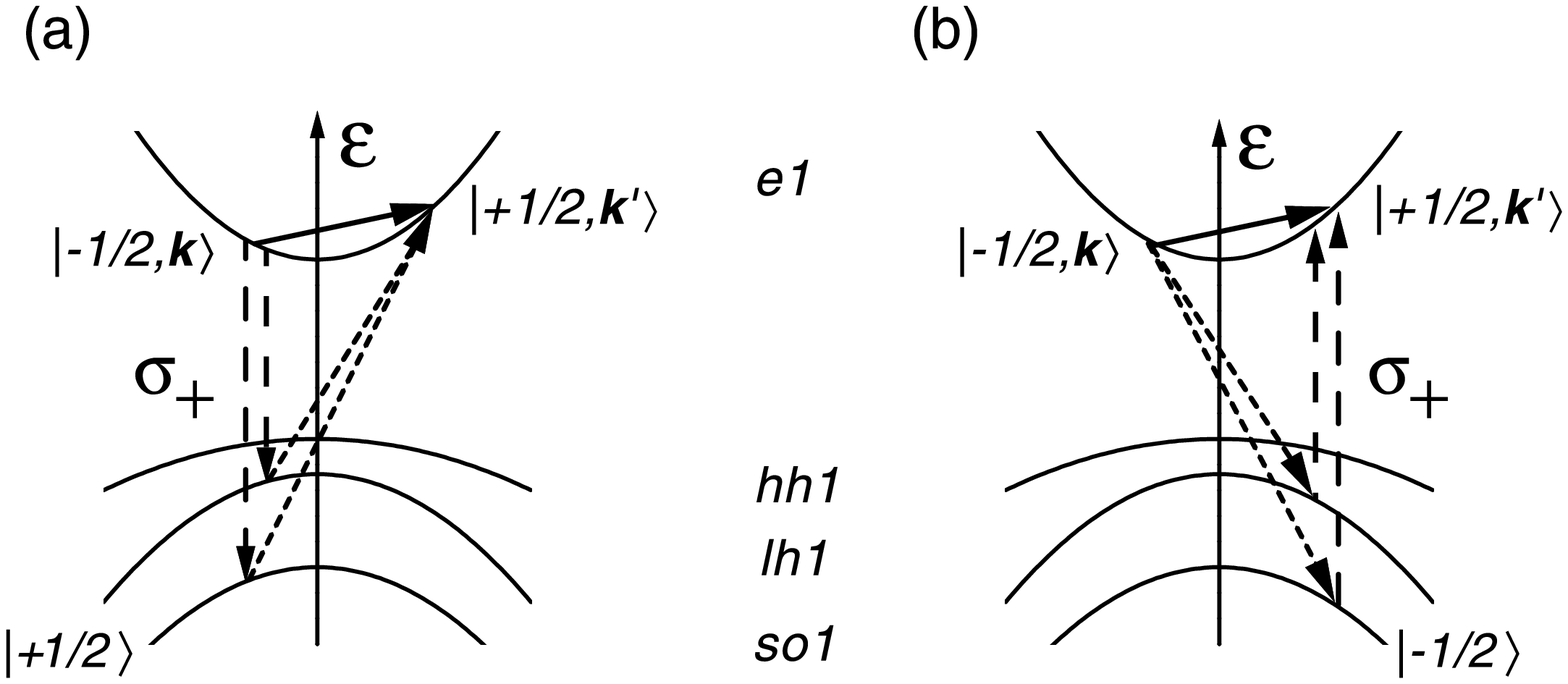 }}
\caption{Sketch of indirect intra-subband optical transitions
(solid arrows) with intermediate states in the valence band.
Dashed and dotted arrows indicate the electron-photon interaction
and the electron momentum scattering.}
\label{fig22}
\end{figure}

In order to obtain optical orientation due to intra-subband
transitions at normal incidence we consider virtual inter-band
transitions with intermediate states in the valence
band~\cite{ICPS26,PASPS02monop}. Fig.~\ref{fig22} demonstrates
schematically the spin orientation at intra-band absorption of
right handed circularly polarized light ($\sigma^{+}$) at normal
incidence. Because of the dipole selection rules for inter-band
optical transitions, the electron transitions with  spin reversal
from $m_s=-1/2$ to $m_s=+1/2$ are possible via intermediate states
in the light-hole and spin-orbit split subbands, while the
opposite processes, $+1/2 \rightarrow -1/2$ are forbidden. As a
result  spin orientation of electrons occurs. At oblique
incidence, the transitions via heavy-hole subbands also contribute
to optical orientation.

For this particular mechanism of monopolar optical orientation one
can derive the following expression for the spin generated per one
absorbed photon of e.g. right-handed circularly polarized radiation
\begin{equation}
s \propto \frac{V^2_{cv}}{V^2_c} \frac{\hbar\omega \,
\Delta^2_{so}}{E_g(E_g+\Delta_{so})(3E_g+2\Delta_{so})} \:.
\label{equ37}
\end{equation}
Here $V_c$ and $V_{cv}$ are the intra-subband and inter-band
matrix elements, respectively, and depend on the mechanism of
momentum scattering. Acoustical-phonon-assisted  and static
impurities processes are considered in~\cite{ICPS26,PASPS02monop}.

\subsection{Spin controlled nonlinearity of spin orientation induced CPGE}
\label{IVD}

Here we discuss the nonlinear behaviour of the spin polarization
induced CPGE. It was observed in \cite{PhysicaE01,PRL02} that the
photocurrent saturates with increasing of the light intensity. In
fact, in this case, the photogalvanic current normalized by the
radiation intensity $I$ is proportional to the
absorbance~\cite{PhysicaE02} and reflects the  power dependence of
the absorption coefficient.

%
\begin{figure}
\centerline{\epsfxsize 120mm \epsfbox{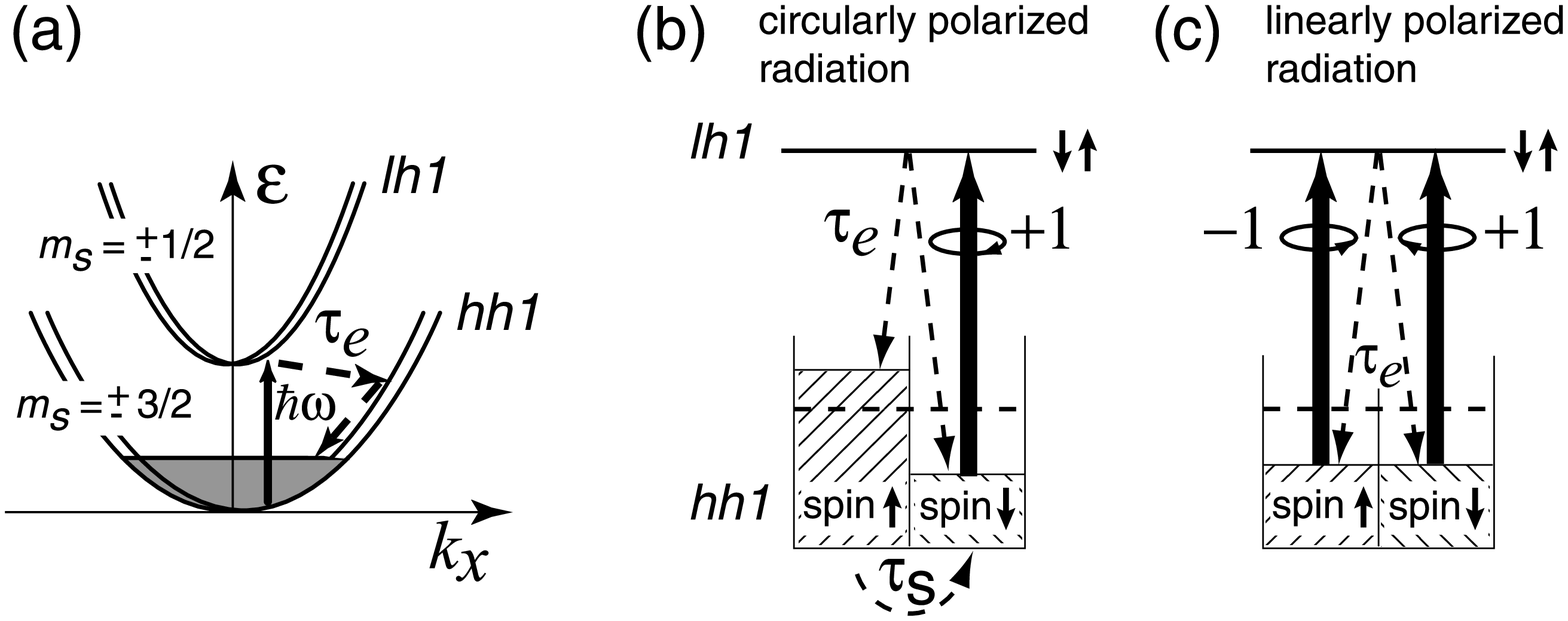  }}
\caption{ Microscopic picture of spin-sensitive bleaching: (a)
sketch of direct optical transitions (full line) between $hh${\it
1} and $lh${\it 1} in $p$-type GaAs/AlGaAs QWs. (b) and (c)
sketches the process of bleaching for   circular and linear
polarized radiation, respectively. Dashed arrows indicate energy
($\tau_e$) and spin ($\tau_s$) relaxation. }
\label{fig23}
\end{figure}

The saturation effect was investigated on $p$-doped QW structures
at direct inter-subband optical transitions. The basic physics of
spin sensitive bleaching of absorption is sketched in
Fig.~\ref{fig23}. Excitation with FIR radiation results in direct
transitions between heavy-hole $hh${\it 1}  and light-hole
$lh${\it 1} subbands. This process depopulates and populates
selectively spin states in $hh${\it 1} and $lh${\it 1} subbands.
The absorption is proportional to the difference of populations of
the initial and final states. At high intensities the absorption
decreases since the photoexcitation rate becomes comparable to the
non-radiative relaxation rate to the initial state. Due to
selection rules only one type of spins is involved in the
absorption of circularly polarized light. Thus the absorption
bleaching of circularly polarized radiation is governed by energy
relaxation of photoexcited carriers and  spin relaxation in the
initial state~(see Fig.~\ref{fig23}b). These processes are
characterized by  energy and  spin relaxation times $\tau_e$ and
$\tau_s$, respectively. We note, that during energy relaxation to
the initial state in  $hh{\it 1}$ the holes lose their
photoinduced orientation due to rapid
relaxation~\cite{Ferreira91p9687}. Thus, spin orientation occurs
in the initial subband $hh{\it 1}$, only. In contrast to
circularly polarized light, absorption of linearly polarized light
is not spin selective and the saturation is controlled by the
energy relaxation only~(see Fig.~\ref{fig23}c). If $\tau_s$ is
larger than $\tau_e$ bleaching of absorption becomes spin
sensitive and the saturation intensity of circularly polarized
radiation drops below the value of linear polarization.

%
\begin{figure}
\centerline{\epsfxsize 86mm \epsfbox{ 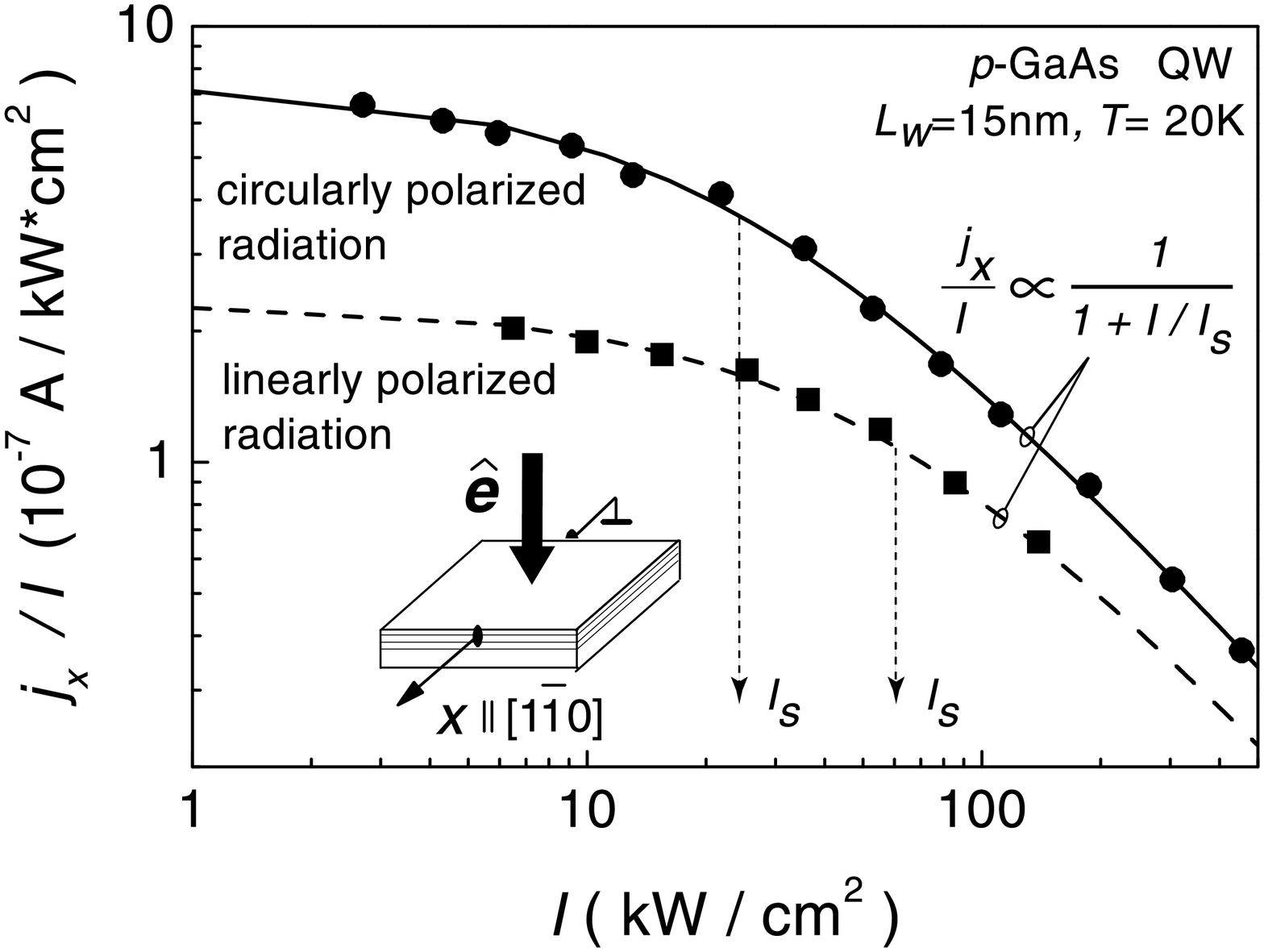 }}
\caption{CPGE and LPGE  currents $j_x$   normalized by intensity
$I$ as a function of $I$ for  circularly and linearly polarized
radiation of $\lambda=148$~$\mu$m, respectively. The  inset  shows
the geometry of the experiment. The current $j_{x }$ is measured
along [1$\bar{1}$0] at normal incidence of radiation on $p$-type
(113)A-grown GaAs QW with $L_W=15$~nm at $T=~$20~K. LPGE was
obtained with the electric field vector {\boldmath$E$} oriented at
45$^\circ$  to the $x$-direction (see Eqs.~\protect(\ref{equ43}))
was used. The measurements  are fitted to $j_x/I\propto
1/(I+I/I_s)$ with one parameter $I_s$ for each state of
polarization (full line: circular, broken line: linear).}
\label{fig24}
\end{figure}

\begin{figure}
\centerline{\epsfxsize 56mm \epsfbox{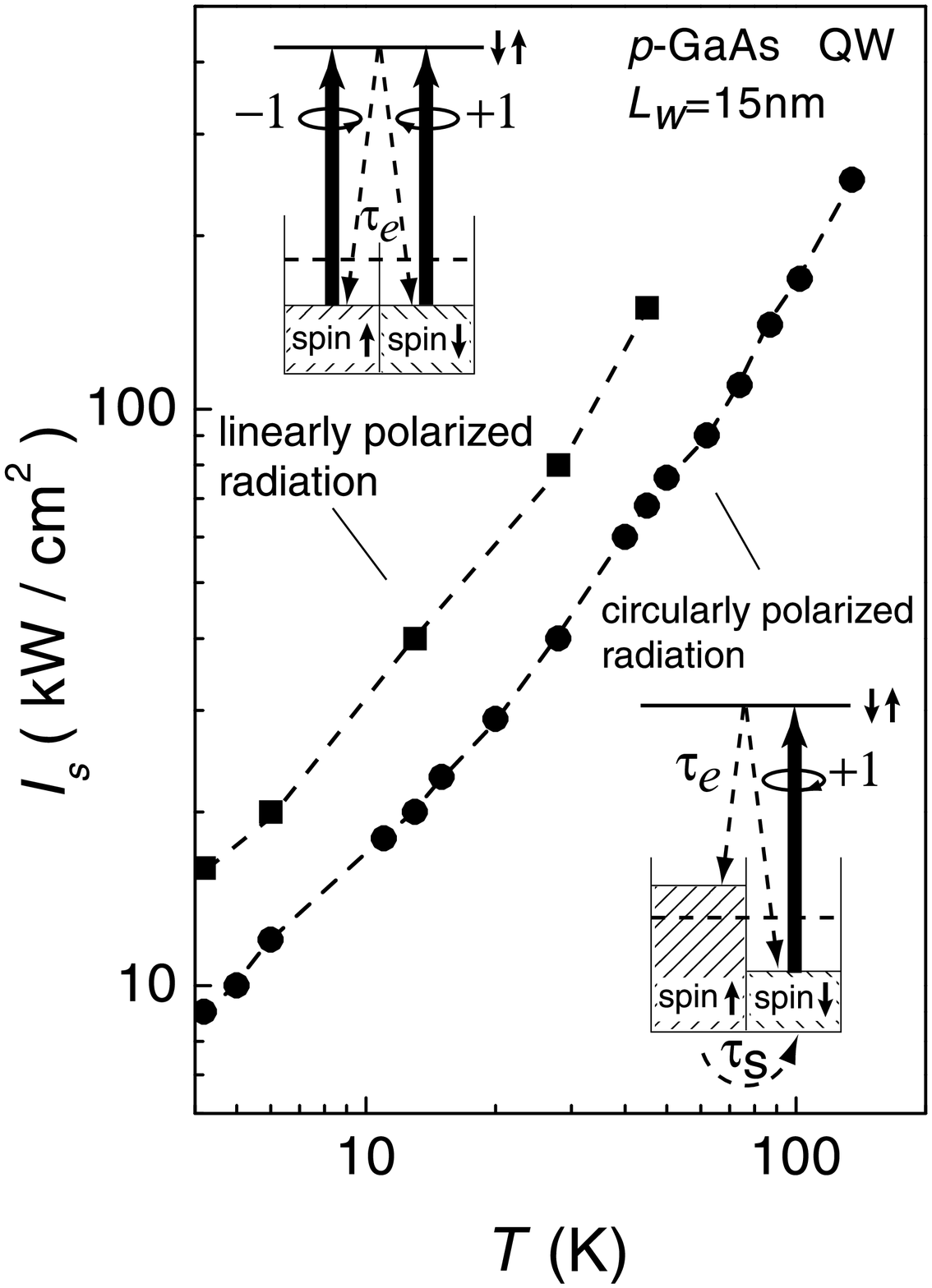  }}
\caption{Temperature dependence of the saturation intensity $I_s$
for linearly and circularly polarized radiation of $\lambda=148
\mu$m. The dependence is shown for a $p$-type GaAs/AlGaAs
(113)A-grown sample with a single QW of $L_W$=15~nm width. The
free carrier density and the mobility were
$1.66\cdot10^{11}$~cm$^{-2}$ and 6.5$\cdot$10$^5$~cm$^2$/(Vs),
respectively.}
\label{fig25}
\end{figure}

\begin{figure}
\centerline{\epsfxsize 86mm \epsfbox{ 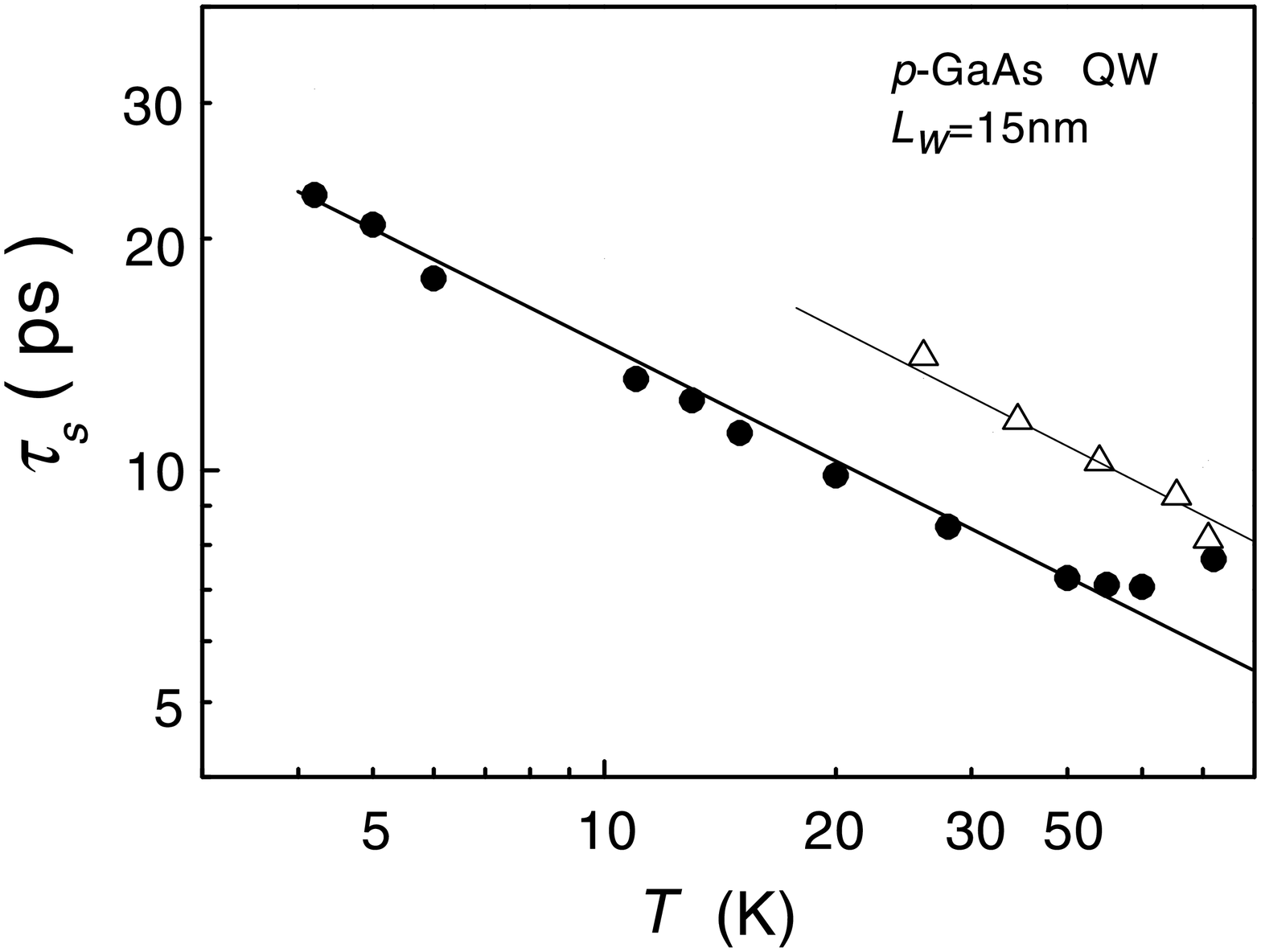 }}
\caption{Experimentally determined spin relaxation times $\tau_s$
of holes in $p$-type GaAs/AlGaAs QWs as a function of temperature
$T$. Open triangles and full dots correspond to (113)A-grown 15~nm
single and multiple (20) QWs, respectively. Free carrier densities
of all samples were about $2\cdot 10^{11}$ cm$^{-2}$ for each QW.}
\label{fig26}
\end{figure}

The difference in absorption bleaching for circularly and linearly
polarized radiation has been  observed experimentally employing
the  spin orientation induced CPGE (see section~2.2) and the
linear photogalvanic effect (see section~\ref{V1})~\cite{PRL02}.
Fig.~\ref{fig24} shows that the photocurrent $j_x$ measured on
$p$-type GaAs QWs depends on the intensity $I$ as $j_x\propto
I/(1+I/I_s)$, where $I_s$ is the saturation intensity. It has been
shown that saturation intensities $I_s$ for circularly polarized
radiation are generally smaller than that for linearly polarized
radiation (Fig.~\ref{fig25}). The non-linear behaviour of
photogalvanic current has been analyzed in terms of
excitation-relaxation kinetics taking into account both optical
excitation and non-radiative relaxation processes. It has been
shown that the photocurrent $j_{LPGE}$ induced by linearly
polarized radiation is described by $j_{LPGE}/I \propto (1 +
I/I_{se})^{-1}$, where $I_{se}$ is the saturation intensity
controlled by energy relaxation of the hole gas. The photocurrent
$j_{CPGE}$ induced by  circularly polarized radiation is
proportional to $(1 + I \left( \frac{1}{I_{se}} + \frac{1}{I_{ss}}
\right))^{-1}$ where $I_{ss}= \hbar\omega p_s /(\alpha_0 d
\tau_s)$ is the saturation intensity controlled by hole spin
relaxation and $\alpha_0$ is the absorption coefficient at low
intensities. Using experimentally obtained $I_{ss}$ together with
the  absorption coefficient $\alpha_0$, calculated
after~\cite{Vorobjev96p981}, spin relaxation times $\tau_s$ have
been derived~(see Fig.~\ref{fig26})~\cite{PRL02}. We note that in
the definition of $I_{ss}$ it was assumed that the spin selection
rule are fully satisfied at the transition energy. This is the
case for optical transitions occurring close to {\boldmath$k$} =
0~\cite{Ferreira91p9687} being realized in the above experiment.
If this is not the case the mixture of heavy-hole and light-hole
subbands reduces the strength of the selection
rules~\cite{Ivchenko96p5852} and therefore reduces the efficiency
of spin orientation. The mixing yields  a multiplicative factor in
$I_{ss}$ increasing the saturation intensity at constant spin
relaxation time~\cite{PhysicaE01}. In~\cite{NOEKS03} a substantial
lowering of $I_{ss}$ with decreasing  QW width  was  observed.
This observation may indicate much slower spin relaxation times
for holes in  narrow QWs as obtained theoretically
in~\cite{Ferreira91p9687}.

%

This result shows that spin relaxation of holes may be obtained by
investigation of the circular photogalvanic effect as a function
of radiation intensity. A substantial portion of  investigations
of the spin lifetime in semiconductor devices are based on optical
spin orientation by inter-band excitation and further  tracing the
kinetics of polarized
photoluminescence
(for review see~\cite{spintronicbook02,Meier,Shah99p243,Vina99p5929}).
These studies give important insights into  the mechanisms of spin
relaxation of photoexcited free carriers. In contrast to these
methods of optical spin orientation,  monopolar spin
orientation allows to study spin relaxation without electron-hole
interaction and exciton formation in the conditions close to the
case of electric spin injection~\cite{Nature02,PRL02,MRS01monopolar}.
Spin photocurrents provide an experimental access to such investigations.

%

\section{Spin-independent photocurrents at homogeneous excitation}
\label{V}

Investigating   spin photocurrents one should take into account
that optical excitation may generate other currents which are not
the result of spin orientation. Spin photocurrents can be
recognized by their dependence on the helicity of the exciting
radiation. Indeed, only spin photocurrents change their direction
 when the state of polarization of radiation is
switched from right- to left-handed or vice versa. This allows to
extract the spin photocurrent contribution from total
photocurrents at visible excitation when spin-independent
photocurrents, like in the case of Dember-effect or photo-voltaic
effect at contacts etc., may occur, being substantially larger than spin
photocurrents. In this situation the periodic modulation of the
state of polarization and synchronous detection of the current is
a convenient technique to measure spin photocurrents by itself. In
the infrared spectral range these strong spin-independent
photocurrents do not occur. However, there are two other sources
of photocurrents at homogeneous excitation  which occur
simultaneously and may be  of the same order of magnitude as spin
photocurrents but again do not require spin orientation. These are
the linear photogalvanic effect (LPGE)~\cite{sturman,book} and the
photon drag effect~\cite{Yaroshetskii80p173,Gibson80p182}.
These photocurrents are not changed in sign or amplitude if the
polarization is switched from $\sigma_+$ to $\sigma_-$.
Furthermore both effects, LPGE and photon drag effect, depend on
the symmetry of the material in a different way than spin
photocurrents. Therefore in most cases it is possible to choose a
crystallographic orientation and an experimental geometry where
only spin photocurrents occur. In other cases the helicity
dependence of  photocurrents allows to distinguish between spin
photocurrents and spin-independent photocurrents.  Based on
phenomenology we discuss spin-independent
photocurrents in  the following and present   some experimental examples.

By general symmetry arguments it can be shown that
these spin-independent photocurrents are given by
\begin{equation}
 j_{\lambda} = \sum_{\mu\nu} \chi_{\lambda \mu \nu} (E_{\mu}
E^*_{\nu} + E_{\nu} E^*_{\mu})/2  + \sum_{\delta\mu\nu}
T_{\lambda\delta \mu \nu }q_{\delta} E_{\mu} E^*_{\nu} \:.
\label{equ38}
\end{equation}
where $\chi_{\lambda\mu\nu}$  and $T_{\lambda\delta \mu \nu }$ are
components of  a third rank and a fourth rank tensor,
respectively, and $q_{\delta}$ is the wavevector of the radiation in
the sample.

The first term on the right-hand side of Eq.~(\ref{equ38}) is
called LPGE because it is independent on the sign of circular
polarization and is usually measured by linearly polarized
radiation. The second term represents  the photon drag effect
which yields a current due to momentum transfer from photons to
electrons.

\subsection{Linear photogalvanic effect} \label{V1}

LPGE arises in homogeneous samples under homogeneous excitation
due to an asymmetry of the scattering of free carriers  on
phonons, static defects, or other carriers in non-centrosymmetric
media (for review see ~\cite{sturman,book}). The linear
photogalvanic effect was observed in some insulators as early as
the 1950th, but was correctly identified as a new phenomenon only
in~\cite{Glass74p233} and theoretically treated
in~\cite{Belinicher77p362}. In bulk materials like GaAs LPGE was
studied in great detailed (for review see
~\cite{Ivchenko80p427,Belinicher80p199,sturman,book}). It has also
been considered and observed in low dimensional structures like
MOSFET Si-structures~\cite{Magaril82p74,Gusev87p33}, III-V
QWs~\cite{APL00,PhysicaE02,Magarill01p652,Schneider98p1289} and
asymmetric SiGe QWs~\cite{PRB02}. Microscopically, the LPGE
current consists of a so-called ballistic and a shift
contribution~\cite{Baltz80p364,Belinicher82p359,Ivchenko84p55}.
The LPGE was successfully applied as a fast detector of the degree
of circular polarization~\cite{Andrianov88p580,Schoenbein95p973}.

As shown in Eq.~(\ref{equ38}) the linear photogalvanic current is
linked to the symmetrized product $E_\mu E^*_\nu$ by a third rank
tensor  $\chi_{\lambda\mu\nu}$ which in turn is symmetric in the
last two indices. Therefore $\chi_{\lambda\mu\nu}$ is isomorphic
to the piezoelectric tensor and may have non-zero components in
media lacking a center of symmetry. Note that in contrast to spin
photocurrents gyrotropy is not necessary. In zinc-blende structure
based QWs of $C_{2v}$ symmetry taking into account only the LPGE
term of Eq.~(\ref{equ38}) we get
\begin{equation}
j_{LPGE,x} = \chi_{xxz} \left( E_x E_z^* + E_z E_x^* \right) \:,\: \: \: \: \: \: \: \: \: \: \:
j_{LPGE,y} = \chi_{yyz} \left( E_y E_z^* + E_z E_y^* \right) \:.
\label{equ39}
\end{equation}
In higher symmetry  QWs of  point-group $D_{2d}$ the  coefficients
are linearly dependent and $\chi_{xxz} = - \chi_{yyz}$.
Eqs.~(\ref{equ39}) shows that the LPGE occurs only at oblique
incidence of radiation because a component of the electric field,
$E_z$, along the $z$-axis is required.

We assume now irradiation of a QW of the $C_{2v}$ symmetry with
the plane of incidence parallel to ($yz$). For linearly polarized
light with an angle $\alpha$ between the plane of polarization
defined by the electric field vector and the $x$-coordinate
parallel to $[1\bar{1}0]$  the LPGE current is given by:
\begin{equation}
j_{LPGE,x} =  \chi_{xxz} \hat{e}_y E^2_0 \sin{2 \alpha}\:,\: \: \: \: \: \: \: \: \: \: \:
j_{LPGE,y} =  \left(
\chi_+ + \chi_- \cos{2 \alpha} \right)  \hat{e}_y E^2_0 \:,
\label{equ40}
\end{equation}
where $\chi_{\pm} = (\chi_{xxz} \pm \chi_{yyz})/2$. In the
experimental setup used for the measurement of the helicity
dependence of spin photocurrents  the laser light is linearly
polarized along $x$ and a $\lambda/4$ plate is placed between the
laser and the sample, Eqs.~(\ref{equ39}) take the form
\begin{equation}
j_{LPGE,x} =
\chi_{xxz} \hat{e}_y E^2_0 \cos{2 \varphi}
\sin{2 \varphi} \:,\: \: \: \: \: \: \: \: \: \: \: j_{LPGE,y} = \left( \chi_+ + \chi_- \cos{2
\varphi} \right) \hat{e}_y E^2_0  \:,
\label{equ41}
\end{equation}

In addition to Eqs.~(\ref{equ39}) the point group $C_s$ allows an
LPGE current at  normal incidence of the radiation because in this
case the tensor {\boldmath$\chi$}
 has the additional non-zero components
$\chi_{xxy^\prime} = \chi_{xy^\prime x}$, $\chi_{y^\prime xx}$
and $\chi_{y^\prime y^\prime y^\prime}$.
This current is given by
\begin{equation}
j_{LPGE,x} =
\chi_{xxy^\prime} \left( E_x E_{y^\prime}^* + E_{y^\prime} E_x^* \right)
\:,\: \: \: \: \: \: \: \: \: \: \:
j_{LPGE,y^\prime} = \left( \chi_{y^\prime xx} |E_x|^2 + \chi_{y^\prime y^\prime y^\prime} |E_{y^\prime}|^2
\right) \:
\label{equ42}
\end{equation}
yielding for linearly polarized light
\begin{equation}
j_{LPGE,x} = \chi_{xxy^\prime}  \hat{e}_{z^\prime} E^2_0   \sin{2
\alpha}\:,\: \: \: \: \: \: \: \: \: \: \:  j_{LPGE,y^\prime} =
\left( \chi_{+}^{\prime} + \chi_{-}^{\prime}  \cos{2 \alpha}
\right) \hat{e}_{z^\prime} E^2_0  \:, \label{equ43}
\end{equation}
where $\chi_{\pm}^{\prime} = (\chi_{xxx} \pm \chi_{xy^\prime y^\prime})/2$ and for the
experimental arrangement of helicity dependence measurements
\begin{equation}
j_{LPGE,x} =
\chi_{xxy^\prime}  \hat{e}_{z^\prime} E^2_0 \cos{2 \varphi}
\sin{2 \varphi}\:,\: \: \: \: \: \: \: \: \: \: \:
j_{LPGE,y^\prime} = \left( \chi_{+}^{\prime} + \chi_{-}^{\prime}
\cos{2 \varphi} \right) \hat{e}_{z^\prime} E^2_0  \:,
\label{equ44}
\end{equation}
The dependence of the LPGE current on the angle of incidence
$\Theta_0$ is determined by the projections of the unit vector
{\boldmath $\hat{e}$} which, as in the case of CPGE, have the form
of Eqs.~(16)-~(18).

Eqs.~(\ref{equ41}) and (\ref{equ44}) for LPGE  at excitation with
elliptically polarized radiation show that it may occur
simultaneously with spin orientation induced CPGE (see Eqs.~(14)
and (15)). However, in this case  LPGE is equal to zero for
circularly polarized radiation for all considered symmetries.
Indeed, light is circularly polarized at $2\varphi = \pi /2, 3\pi
/2 \dots$ therefore for $\sigma_{\pm}$ radiation $\cos 2\varphi$
in Eqs.~(\ref{equ41}) and (\ref{equ44}) is equal to zero and the
LPGE current vanishes. For $P_{circ}$ between -1 and +1, as in the
measurements  of helicity dependence of the  photocurrent both,
spin orientation induced CPGE and LPGE can be simultaneously
present. However, measurements carried out on
GaAs~\cite{APL00,PRL01,PhysicaE02,ICPS26invited},
InAs~\cite{PRL01,PhysicaE02}, and ZnSeMnTe~\cite{DPG01} QWs in the
whole wavelength range as well as on SiGe QWs in the mid
infrared~\cite{PRB02,PASPS02sige} have shown that the contribution
of the LPGE is negligible (see Figs.~10 and~13).

%
\begin{figure}
\centerline{\epsfxsize 86mm \epsfbox{ 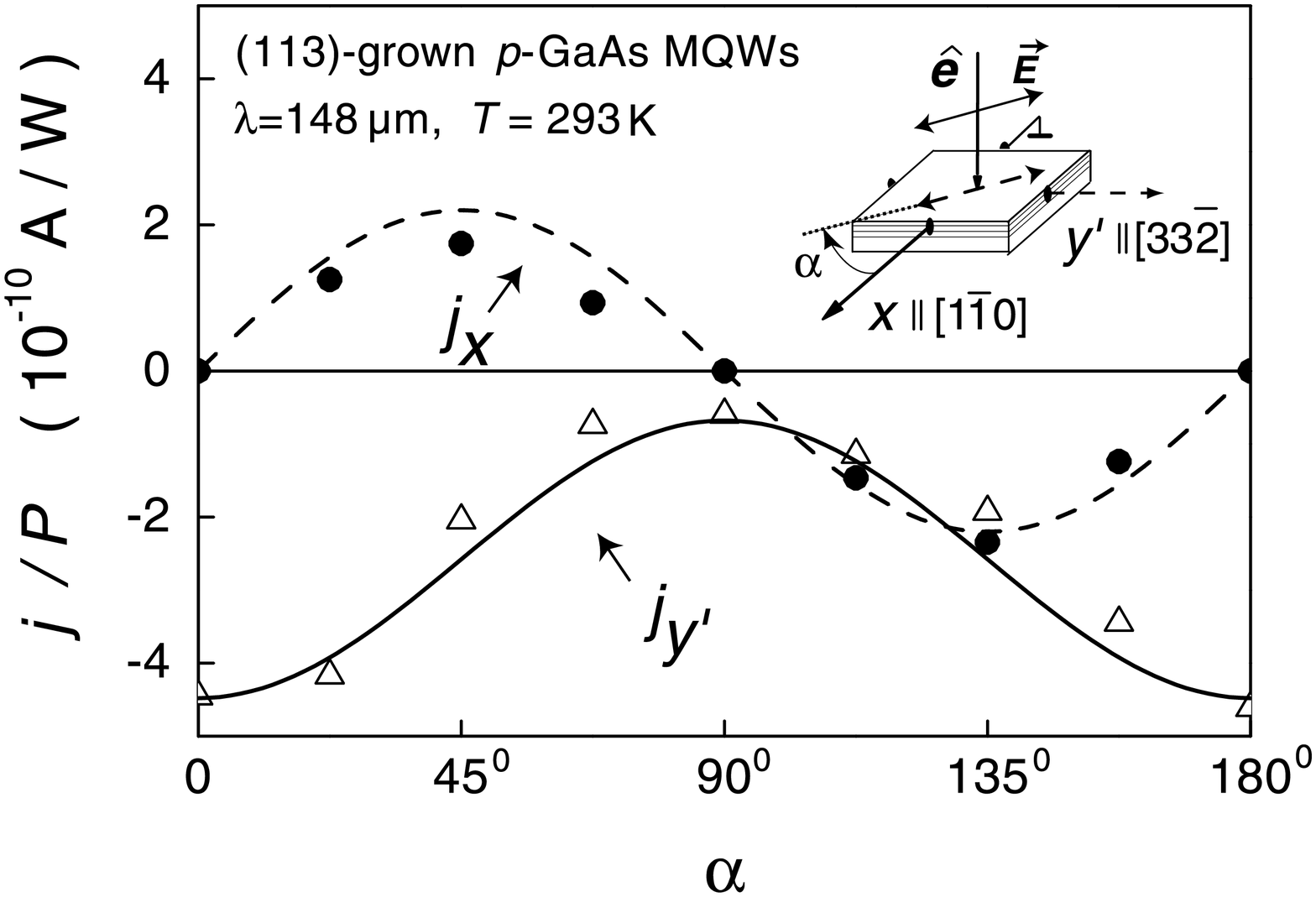 }}
\caption{Linear photogalvanic current  $j$ normalized by $P$ as a
function of the angle $\alpha$ between the plane of linear
polarization and the axis $x$. Data are obtained for $x$ and
$y^\prime$ direction under normal incidence. The broken line and
the full line are fitted after Eqs.~\protect (\ref{equ43}). }
\label{fig27}
\end{figure}

The situation is different for SiGe QWs at the FIR excitation
where spin orientation induced CPGE and LPGE with comparable
strength have been detected~\cite{PRB02,PASPS02sige}. Fig.~15
exhibits  experimental data obtained on $p$-type SiGe (113)-grown
QW structure. Broken and dotted lines show contributions of the
circular photocurrent, $j_x \propto \sin{2 \varphi}$, and the
linear photocurrent, $j_x \propto \sin{2 \varphi}\cdot\cos{2
\varphi}$, respectively. The tensors {{\boldmath$\gamma$} and
{\boldmath$\chi$}  describe different physical mechanisms and,
therefore, may depend differently on the material parameters,
excitation wavelength, and temperature. Obviously in most cases
the contribution of {{\boldmath$\gamma$} to the photocurrent is
larger than that of {{\boldmath$\chi$}.

%
\begin{figure}
\centerline{\epsfxsize 86mm \epsfbox{ 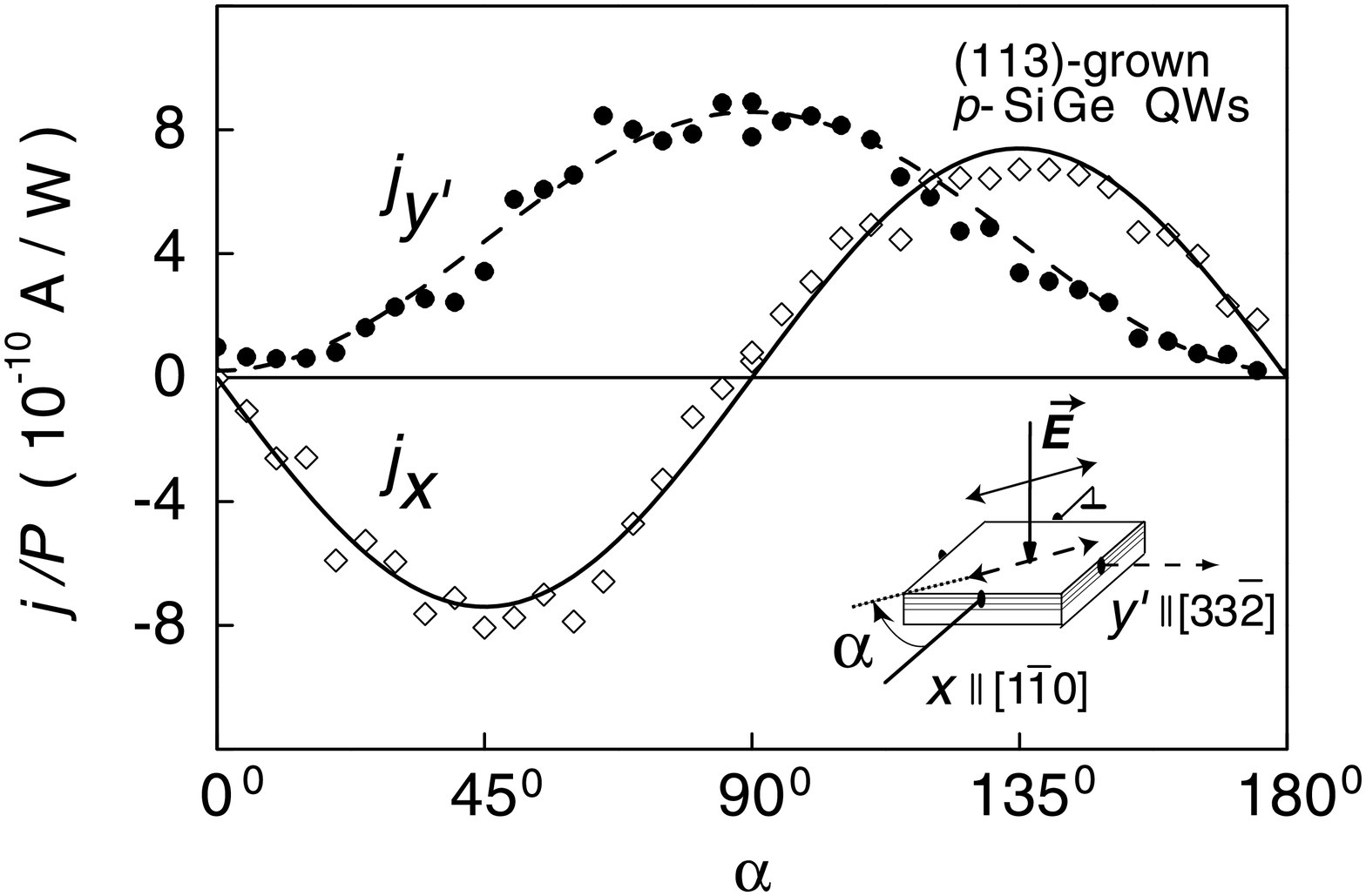 }}
\caption{Linear photogalvanic current $j$ normalized by $P$ for
$x$ and $y^\prime$ direction as a function of the angle $\alpha$
between the plane of linear polarization and the axis $x$. The
results were obtained  at room temperature with  (113)-grown SiGe
QWs under normal incidence of irradiation at  $\lambda =
280~\mu$m. The broken line and the full line are fitted after
Eqs.~\protect (\ref{equ43}).}
\label{fig28}
\end{figure}

The occurrence of LPGE without CPGE is demonstrated by
measurements at linear polarization at which  CPGE is zero,
depicted in Fig.~\ref{fig27} for GaAs QWs and in Fig.~\ref{fig28}
for SiGe QWs. It was also observed at the excitation by elliptical
polarized radiation with $P_{circ}$ varying from -1 to +1 in a
geometrical direction where CPGE is forbidden by symmetry. This is
demonstrated in Fig.~\ref{fig29} where a longitudinal current
generated by oblique incidence along [110] direction is shown.

%
\begin{figure}
\centerline{\epsfxsize 86mm \epsfbox{ 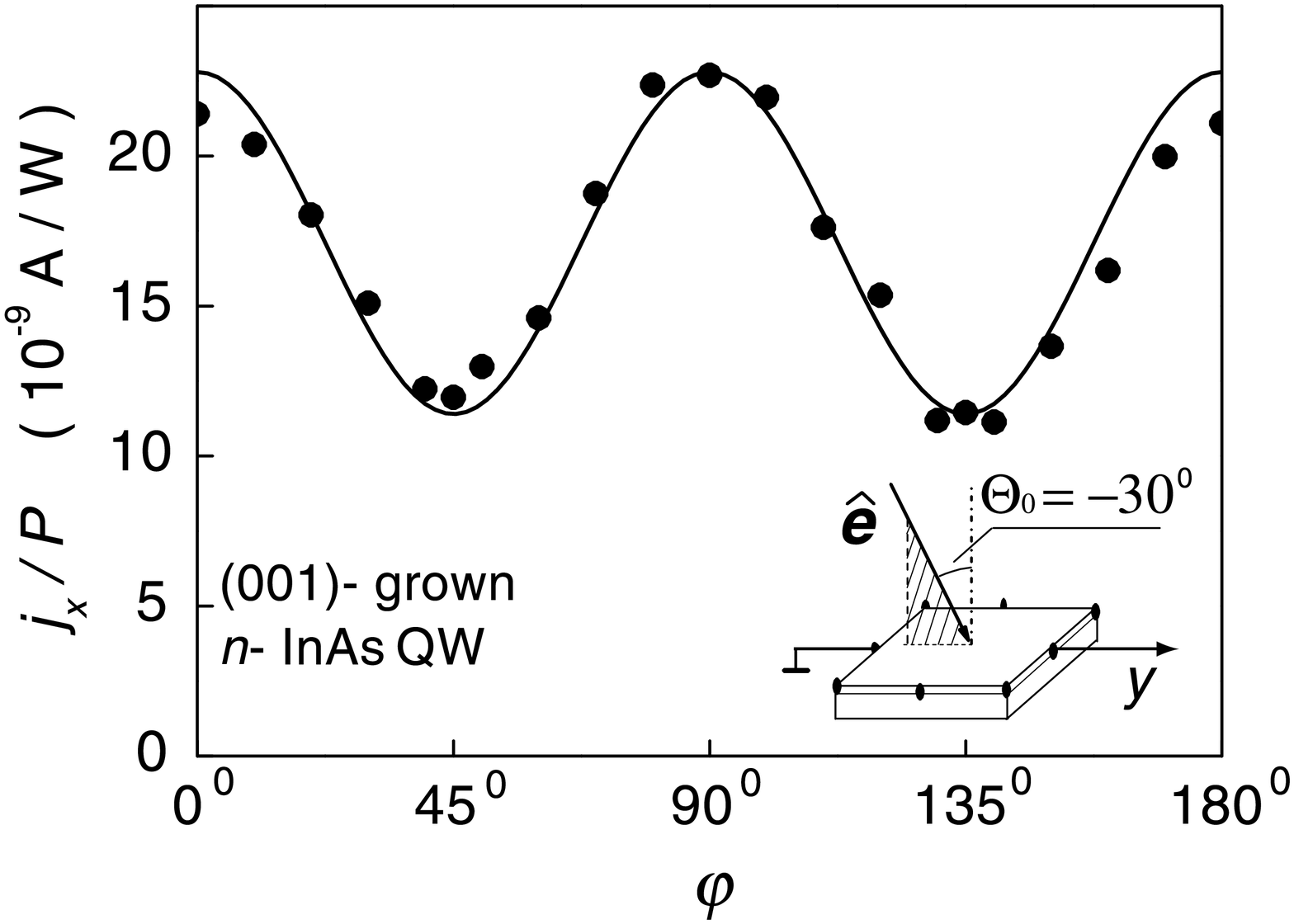 }}
\caption{Linear photogalvanic current $j_{y}$  normalized by $P$
as a function of the phase angle $\varphi$. The solid curve is
fitted after Eq.~\protect(\ref{equ41}). }
\label{fig29}
\end{figure}

\subsection{Photon drag effect} \label{V2}

The photon drag effect arises due to a momentum transfer from
photons to free carriers. This effect was first observed in bulk
semiconductors~\cite{Danishevskii70p544,Gibson70p75} and  has been
investigated in various materials like Ge, Si and  GaP in a wide
range of optical excitation mechanisms as  inter-band transitions,
direct and indirect transitions at free carrier absorption, at
impurity ionization etc. (for review
see~\cite{Yaroshetskii80p173,Gibson80p182}). It was also
intensively studied  in GaAs and InAs
QWs~\cite{PhysicaE02,Luryi87p2263,Wieck90p463,Grinberg88p87,Dmitriev91p462,Keller93p4786,Beregulin94p85,Vasko96p9576,Vasko98p15666}.
The photon drag is of great technical importance for fast infrared
and far-infrared detection of short laser
pulses~\cite{Ganichev84p20,Beregulin90p853,Yaroshetskii80p173,Gibson80p182,Sigg95p2827}.

The photon drag current described by the second term of the right
hand side of Eq.~(\ref{equ38}) is mediated by a fourth rank tensor
{\boldmath$T$}. Therefore there is no symmetry  restriction for
this effect. In $T_{\lambda\delta\mu\nu}$ the first index
$\lambda$ runs over $x$ and $y$ only because the current must be
confined in the plane of the QW.

For QWs of  $C_{2v}$ symmetry
the photon drag effect yields the current
\begin{eqnarray}\label{equ45}
j_{PD,x} =  \sum_{\mu=x,y,z} T_{x x\mu \mu } q_x |E_{\mu}|^2
+ T_{xyxy}q_y  \left( E_x E_y^* + E_y E_x^* \right) \:,\\
j_{PD,y} =  \sum_{\mu=x,y,z} T_{y y\mu \mu }q_y  |E_{\mu}|^2
+ T_{yxyx} q_x \left( E_x E_y^* + E_y E_x^*\right)
\:.\nonumber
\end{eqnarray}
For $C_s$ symmetry this equation has the same form if $y$ and $z$ are replaced by the primed coordinates.
In higher symmetric QWs of the point group $D_{2d}$ the number of independent
non-zero tensor components is reduced by
\begin{equation} \label{equ46}
T_{xxxx} = T_{yyyy} \:,\: \: \: \: T_{xxyy} = T_{yyxx} \:,\: \: \: \:  T_{xxzz} = T_{yyzz} \:,\: \: \: \:  T_{xyxy} = T_{yxyx}\:.
\end{equation}
The above equations show that the photon drag effect occurs in QWs
of all symmetries at oblique incidence of radiation only. The
longitudinal effect, first term on the right hand side of
Eqs.~(\ref{equ45}), is usually much stronger than the transverse
effect described by the second term. This is the reason that in
the transverse geometry of spin orientation induced CPGE no
influence of the photon drag effect could be detected as yet.
However, in the longitudinal geometry the photon drag effect
yields a measurable current. Again helicity dependence can help to
distinguish spin photocurrents from the photon drag effect. On the
other hand the separation of LPGE and photon drag effect is not so
obvious. The usual method to identify photon drag effect is based
on the sign inversion of the current by reversing  the wavevector
of light in the plane of the sample. The same sign inversion
occurs also  for LPGE (see Eqs.~(16), (18), (\ref{equ40}), and
(\ref{equ41})).

Both effects may be distinguished from polarization dependencies.
The situation is simple for transversal currents. Indeed, LPGE
vanishes for linearly polarized radiation with the radiation
electric field normal or parallel to the current flow (see left
equation of Eqs.~(\ref{equ40}),(\ref{equ43})) or  for circularly
polarized radiation (see left equation of
Eqs.~(\ref{equ41}),(\ref{equ44})) whereas a photon drag effect may
be present (see Eqs.~(\ref{equ45})). However, for longitudinal
currents  the photon drag effect and the LPGE may be present at
the same time with comparable strength for any polarization. In
contrast to the transverse effect, longitudinal LPGE  has a
polarization independent term $\chi_+$ in the right equation of
Eqs.~(\ref{equ40}),(\ref{equ41}) and $\chi_+^\prime$ in the right
equation of Eqs.~(\ref{equ43}),(\ref{equ44}). Thus, a longitudinal
current  in non-centrosymmetric QWs which changes sign at reversal
of light propagation need not to be the photon drag current. The
characteristic polarization dependencies as well as the helicity
dependence may help to identify the underlying microscopic
mechanisms.  We note that by investigation of the photon drag
effect in  QWs without inversion center one should $always$ take
into account the LPGE contribution and vice versa. For elliptical
polarization the spin orientation induced CPGE may also contribute
in the total current.

As a concluding remark, spin photocurrents  in any case  can be
distinguished  from helicity independent currents by switching
the helicity from right to left or the other way round. The
fraction of spin photocurrents in the total current can
quantitatively be extracted by modulation methods.

\section{Spin photocurrents caused by inhomogeneities}
\label{VI}

One of the essential features of  spin photocurrents reviewed here
is the homogeneity of both the optical excitation and the
distribution of spin polarization in a two-dimensional electron
gas of gyrotropic QWs. However, spin photocurrents may occur due
to an inhomogeneous spin distribution obtained by inhomogeneous
optical excitation or in bulk inhomogeneities like
$p-n$ junctions.

%
\begin{figure}
\centerline{\epsfxsize 120mm \epsfbox{ 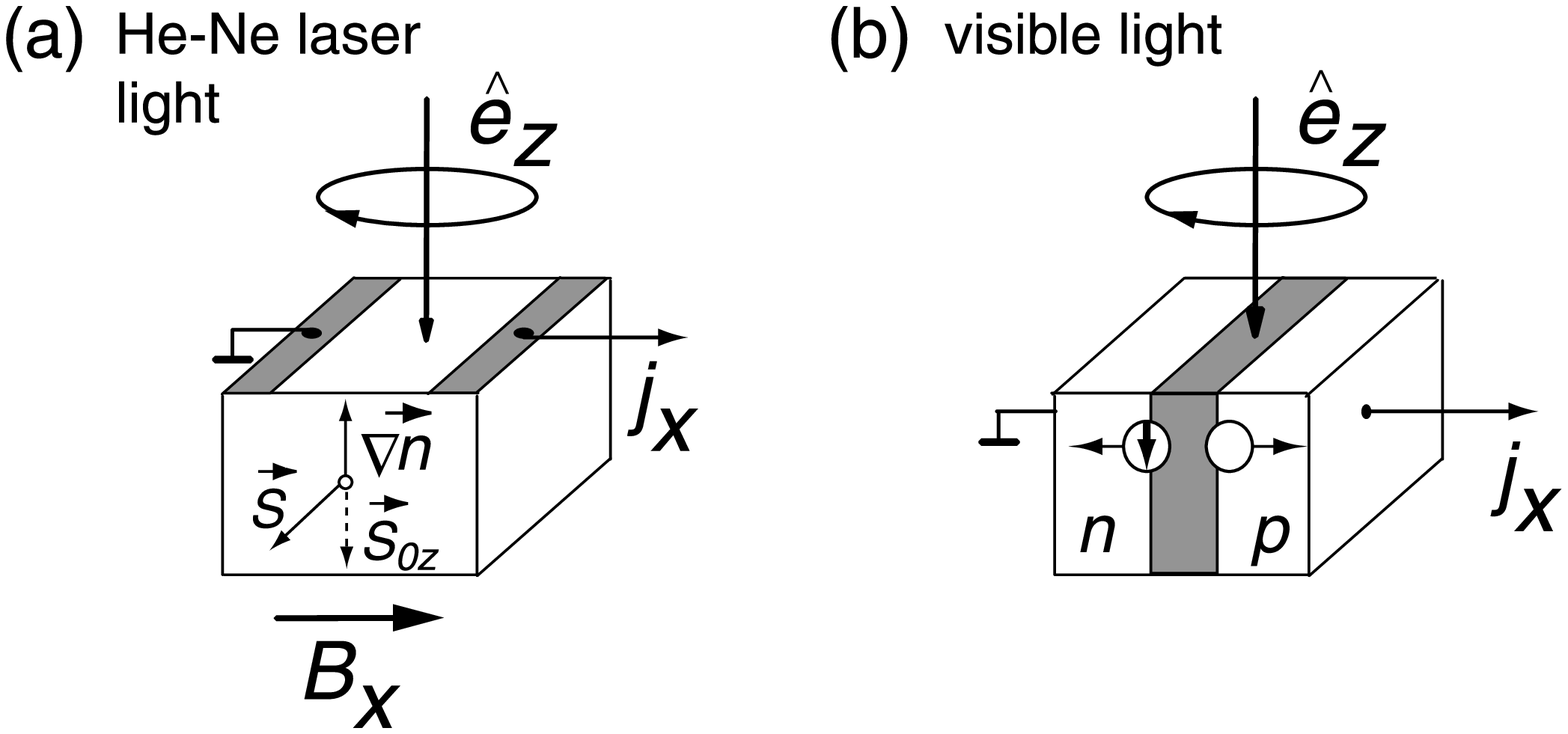}}
\caption{ Sketches of the experimental arrangement  of: (a)
Surface spin photocurrent due to non-homogeneous spin orientation
after~\protect\cite{Bakun84p1293}. $\nabla${\boldmath$n$}
indicates gradient of electron density in the penetration depth of
strong fundamental absorbtion. (b) Spin-voltaic effect after
~\protect\cite{Zutic01p1558,Zutic02p066603}.
 Circle with arrow indicates spin polarized electrons,
 open circle represents unpolarized  holes due to rapid spin relaxation.
}
\label{fig30}
\end{figure}

As it was already introduced a surface current {\boldmath $j$} $=
e\cdot \zeta \cdot rot${\boldmath $S$} due to inhomogeneity of
spin polarization {\boldmath$S$} of electrons at a semiconductor
surface layer proposed in~\cite{Averkiev83p393} was observed in
bulk AlGaAs samples~\cite{Bakun84p1293}. Here $\zeta$ is a
coefficient proportional to the energy of the spin-orbit splitting
of the valence band $\Delta_{so}$. The inhomogeneous spin
polarization was obtained by the strong absorption of circularly
polarized radiation at the band edge of AlGaAs mixed crystals
(HeNe-laser excitation). The radiation was of normal incidence  on
the sample resulting in a gradient of the spin density into the
material due to surface generation and diffusion into the bulk.
Spin-orbit interaction at the surface yields asymmetric electron
scattering which gives rise to a current in the direction
perpendicular to the gradient of spin density and the average
electron spin. An external magnetic field oriented in the surface
plane was used to optimize the current by rotating the spin
polarization into the surface plane (see
Fig.~\ref{fig30}a)~\cite{Bakun84p1293}.  The surface current shows
the Hanle-effect as a function of the magnetic field strength. The
geometry and the experimental procedure are very similar to that
used  to demonstrate the spin-galvanic effect (see Fig.~7). The
crucial difference to the spin-galvanic effect is that in this
case of surface photocurrent caused by optical orientation a
gradient of spin density is needed. Naturally this gradient is
absent in QWs where the spin-galvanic effect has been investigated
because QWs are two-dimensional (no `thickness').

Another type of spin photocurrents recently was  proposed in
analogy to the photo-voltaic effect in $p-n$
junctions~\cite{Zutic01p1558,Zutic02p066603}. This spin-voltaic
effect occurs due to uniform illumination of a $p-n$ junction with
circularly polarized inter-band light resulting in spin
polarization of a charge current. Circular polarization generates
spin polarized electrons and holes. Due to the fast relaxation of
hole spin polarization in the bulk and the long spin lifetime of
electrons, the photocurrent becomes spin polarized. Indeed by the
built-in electric field $E_{bi}$ spin polarized electrons are
swept to the $n$-side and the unpolarized holes drift to the
$p$-side of the junction (see Fig.~\ref{fig30}b).

\section{Summary}
\label{VII}

A non-equilibrium uniform spin polarization obtained by optical
orientation drives an electric current in QWs if they belong to a
gyrotropic crystal class. In QWs prepared from zinc-blende
structure materials gyrotropy is  naturally given due to the lack
of inversion symmetry in the basic material which itself is not
gyrotropic. In QWs based on  diamond structure materials, like Si
and Ge which possess a center of inversion,  gyrotropy may be
introduced by artificially growing of asymmetric structures. In
gyrotropic QWs spin-orbit interaction  results in a spin splitting
in {\boldmath$k$}-space of subbands yielding the basis of spin
photocurrents. Two different microscopic mechanisms of spin
photocurrents can be distinguished, spin orientation induced
circular photogalvanic effect and  spin-galvanic effect. In the
first effect the coupling of the helicity of light to spin
polarized final states with a net linear momentum is caused by
angular momentum selection rules together with band splitting in
{\boldmath$k$}-space due to {\boldmath$k$}-linear terms in the
Hamiltonian. The current flow in the second effect is driven by
asymmetric spin relaxation of a homogeneous non-equilibrium spin
polarization. The current is present even if the initial electron
distribution in each spin-split subband is uniform.

The
experimental results on spin photocurrents due to homogeneous spin
polarization are in good agreement to the phenomenological theory.
Both mechanisms of spin photocurrents as well as the removal of
spin degeneracy in  {\boldmath$k$}-space are described by second
rank pseudo-tensors.  Because of tensor equivalence in each symmetry the
irreducible components of these tensors differ by scalar factors
only.
Therefore macroscopic measurements
of photocurrents in different geometric configurations of
experiments allow to conclude on details of the microscopic
tensorial spin orbit interaction. In particular the relation
between the symmetric and anti-symmetric part of the spin-orbit
interaction representing  Dresselhaus like terms (including
interface inversion asymmetry) and the Rashba term, respectively,
may be obtained. Furthermore the macroscopic symmetry of QWs may
easily be determined.

Spin photocurrents are obtained by circularly polarized radiation.
The most important feature of spin photocurrents is their helicity
dependence. The current is proportional to helicity and reverses
its direction upon changing the handedness of radiation. The
effect is a quite general property of QWs and has already been
observed in many different $n$- and $p$-type semiconductor
structures at various kinds of optical excitation like inter-band
and free carrier absorption. It is present in materials of
different mobilities, also at very low mobility, in a wide range
of carrier densities and can be detected even at room temperature.
Spin photocurrents are not limited to 2D structures. Most recently
they have been  predicted for gyrotropic 1D systems like carbon
nanotubes of spiral symmetry ~\cite{Ivchenko02p155404}. The effect
is caused by  coupling between the electron wavevector along the
tube principal axis and the orbital momentum around the tube
circumference.

Spin photocurrents were applied to investigate the mechanism of
spin relaxation at monopolar spin orientation where only one type
of charge carriers is involved in the excitation-relaxation
process. This condition is close to that of electrical spin
injection in semiconductors. Two methods were applied to determine
spin relaxation times: the Hanle effect in the spin-galvanic
current  and  spin sensitive bleaching of  photogalvanic currents.
The spin orientation induced CPGE has also been applied to detect
the state of polarization of terahertz
radiation~\cite{MRS02detector}. The rapid momentum relaxation at
room temperature in quantum well yields picosecond time
resolution.

\section*{Acknowledgement}
\addcontentsline{toc}{section}{Acknowledgement}

The authors thank  E.L.~Ivchenko, V.V.~Bel'kov,  L.E.~Golub,
S.N.~Danilov and Petra~Schneider for many discussions and helpful
comments on the present manuscript. We are also indebted to
G.~Abstreiter, V.V.~Bel'kov, M.~Bichler, J.~DeBoeck, G.~Borghs,
K.~Brunner, S.N.~Danilov, J.~Eroms, E.L.~Ivchenko, S.~Giglberger,
P. Grabs, L.E.~Golub, T.~Humbs, J.~Kainz, H.~Ketterl, B.N.~Murdin,
Petra~Schneider, D.~Schuh, M.~Sollinger, S.A.~Tarasenko,
L.~Molenkamp, R.~Newmann, V.I.~Perel, C.R.~Pidgeon,
 P.J.~Phillips, U.~R\"{o}ssler, W.~Schoepe, D.~Schowalter, G. Schmidt, V.M.~Ustinov, L.E.~Vorobjev,
 D.~Weiss, W.~Wegscheider, D.R.~Yakovlev, I.N.~Yassievich, and A.E.~Zhukov,  for long
standing cooperation during the work on spin photocurrents. We
gratefully acknowledge financial support by the Deutsche
Forschungsgemeinschaft (DFG), the Russian Foundation for
Basic Research (RFBR) and the NATO Linkage Grant which
made this work possible.




\end{document}